\documentclass[%
 aip,
 amsmath,amssymb,
 reprint,%
 floatfix,
]{revtex4-2}

\usepackage{graphicx}
\usepackage{dcolumn}
\usepackage{bm}

\usepackage[utf8]{inputenc}
\usepackage{mathptmx}
\usepackage{etoolbox}

\usepackage[version=3]{mhchem} 
\usepackage{xcolor}
\usepackage{multirow}
\usepackage{subfigure}
\usepackage{siunitx}
\usepackage{booktabs}

\makeatletter
\def\@email#1#2{%
 \endgroup
 \patchcmd{\titleblock@produce}
  {\frontmatter@RRAPformat}
  {\frontmatter@RRAPformat{\produce@RRAP{*#1\href{mailto:#2}{#2}}}\frontmatter@RRAPformat}
  {}{}
}%
\makeatother

\newcolumntype{d}[1]{D{.}{.}{#1}}

\DeclareSIUnit{\angstrom}{\text{\r{A}}}

\usepackage[normalem]{ulem}

\begin{document}
\author{Aaron M. Schankler}
\altaffiliation
{These authors contributed equally.}
\affiliation
{Department of Chemistry, University of North Carolina, Chapel Hill, North Carolina 27599, USA}

\author{Ruyi Song}
\altaffiliation
{These authors contributed equally.}
\affiliation
{Department of Chemistry, Duke University, Durham, North Carolina 27708, USA}

\author{Sampreeti Bhattacharya}
\altaffiliation
{These authors contributed equally.}
\affiliation
{Department of Chemistry, University of North Carolina, Chapel Hill, North Carolina 27599, USA}

\author{Ela Lucas}
\affiliation
{School of Engineering, Brown University, Providence, Rhode Island, 02912, USA}

\author{Lee Hampton III}
\affiliation
{Duke University Program in Materials Science and Engineering, Duke University, Durham, North Carolina 27708, USA}

\author{Alexandre Tkatchenko}
\affiliation
{Department of Physics and Materials Science, University of Luxembourg, L-1511 Luxembourg City, Luxembourg}

\author{Yosuke Kanai}
\affiliation
{Department of Chemistry, University of North Carolina, Chapel Hill, North Carolina 27599, USA}
\affiliation
{Department of Physics and Astronomy, University of North Carolina, Chapel Hill, North Carolina 27599, USA}
\email{ykanai@unc.edu}

\author{Volker Blum}
\affiliation
{Department of Chemistry, Duke University, Durham, North Carolina 27708, USA}
\affiliation
{Thomas Lord Department of Mechanical Engineering and Materials Science, Duke University, Durham, North Carolina 27708, USA}
\email{volker.blum@duke.edu}

\title
{{
Updated Tkatchenko Scheffler method and assessment of dispersion methods for alkali-containing systems
}}

\begin{abstract}
The Tkatchenko-Scheffler (TS) pairwise method to calculate dispersion interactions is a widely used approach to incorporate missing long-range van der Waals contributions in semilocal and hybrid density functional calculations.
{Despite subsequent many-body extensions, this formulation remains widely used.}
In 2018, Fedorov \emph{et al.} reported updated van der Waals radii to the seminal work published in 2009.
The present work examines the accuracy of the TS method with updated van der Waals radii (TS\_2018), coupled with the semilocal Perdew-Burke-Ernzerhof density functional, for structural predictions of semiconducting and insulating materials.
Special attention is paid to materials containing alkali elements, for which the TS\_2009 method exhibits a large overbinding, associated with potentially large errors in predicted atomic structures.
We also consider a narrower reformulation (TS\_alkali) where only the the alkali atoms are corrected, so the method remains otherwise compatible with TS\_2009.
{We compare these methods with a broader set of dispersion corrections---the DFT-D3 method, the exchange-hole dipole moment method, and the beyond-pairwise, nonlocal many-body dispersion method---for ionic and alkali-containing systems more generally. We examine equilibrium geometry predictions for}
45 inorganic solid compounds as well as three Cs-containing halide perovskites, CsPb$X_3$ ($X$ = Cl, Br, I).
{Our results show that the updated van der Waals radii in the TS\_2018 and TS\_alkali methods resolve the overbinding that is present with TS\_2009 in alkali-containing systems, bringing predictions into closer agreement with other dispersion methods.
For the halide perovskites, method-dependent differences in predicted cell volumes and shapes persist, and no approach performs with perfect accuracy on all metrics for all three materials.}
\end{abstract}

\maketitle

\section{Introduction}

In materials science and chemistry, van der Waals (vdW) dispersion interactions are frequently far weaker than the primary bonding interactions. \cite{langbein1974theory,klimevs2012perspective}
However, when comparing systems with similar energetics based on their primary bonding structure, dispersion energy can be critical for accurate description of the structural, cohesive, and therefore even electronic properties.
This is especially true in systems where molecular conformation plays a key role\cite{jarvis2015measuring, rossi2014validation, schubert2015exploring}, in systems consisting of spatially separated moieties (including molecular crystals) \cite{firaha2023predicting, bucko2010improved}, or in systems that contain soft, flexible structural degrees of freedom (e.g., halide perovskites and their nanostructures) \cite{gao2019molecular, park2023thickness, jana2020organic, kim2022structural, song2023structure, song2024density}.
In such systems, the absence or presence of vdW interactions can impact computational predictions of simple crystallographic parameters, \cite{french2010long, gao2013electronic, wang2014density, ferjani2022first}
leading to, for example, a \qty{24}{\percent} underestimation of the density of benzene molecular crystals\cite{lu2009abinitio} or a \qty{32}{\percent} overestimation of the inter-layer spacing of graphite.\cite{kim2020umbd}
In molecular crystals, vdW interactions can change the relative energetics of competing conformers,\cite{rossi2014validation,schubert2015exploring}
while in 2D and 3D halide perovskites, they can impact the unit cell.\cite{liu2018tunable, heine2023benchmark}

In this paper, we benchmark {several vdW methods, including} the method of Tkatchenko and Scheffler (TS\_2009) \cite{tkatchenko2009accurate} with the updated parameters proposed by Fedorov \emph{et al.} \cite{fedorov2018quantum} (TS\_2018).
Historically, the TS\_2009 method has been widely used, and it has the advantage of being simple, robust, and well-benchmarked.
However, difficulties with modeling materials containing alkali ions using this and other vdW methods have been identified in the literature.\cite{blowey2020alkali, kim2022structural, kim2020umbd}
These drawbacks are addressed in the TS\_2018 method, so we undertake a benchmark to characterize the relationship between this method and many existing results in the literature using the TS\_2009 method.
We also describe a more targeted modification to the TS\_2009 method (TS\_alkali) that retains the accuracy improvements on compounds containing alkali ions while otherwise maintaining backwards compatibility with the TS\_2009 method.
We note that a host of other vdW methods exist in the literature
(including the DFT-D methods of Grimme \emph{et al.} \cite{grimme2004accurate}, the many-body dispersion method of Tkatchenko \emph{et al.} \cite{tkatchenko2012accurate, ambrosetti2014long, hermann2020density}, the exchange-hole dipole moment (XDM) model \cite{becke2007exchange,rumson2023low, price2023xdm}, and others)
that each have attractive properties, and in some cases claim improvement over the TS\_2009 method.
For the present work,
{we additionally compare these results with benchmarks of the DFT-D3 method with Becke-Johnson damping\cite{grimme2010consistent, grimme2011effect}, the XDM method coupled with the B86bPBE semilocal density functional\cite{oterodelaroza2020application, price2023xdm}, and the nonlocal manybody dispersion method\cite{hermann2020density}.}
{These comparisons assess whether the corrected alkali vdW radii bring the updated TS methods into agreement with other dispersion approaches across different bonding environments and } establish the benchmark foundation needed to solidly ground their future use, e.g., for organic-inorganic hybrids, especially those containing alkali ions.

Density functional theory (DFT) is arguably the most important approach for atomistic simulations of materials and molecules today. It became clear very early that some of the most practical density functional approximations (DFAs), i.e., semilocal and hybrid DFAs, do not account for nonlocal correlation terms that give rise to dispersion interactions \cite{kristyan1994can, wu2002empirical, grimme2004importance, grimme2006semiempirical}
High-level wavefunction methods such as coupled cluster theory do include this long-range correlation, but their computational cost is is significantly higher than that of semilocal and hybrid DFT for extended systems\cite{szabo2012modern}.
The vdW term can also be calculated directly from the adiabatic-connection fluctuation-dissipation (ACFD) theorem using the random phase approximation (RPA) \cite{bohm1951collective, fuchs2002accurate, stohr2019theory}
{and the related renormalized second order perturbation theory (rPT2), which also includes second order screened exchange and renormalized single excitation terms\cite{ren2013renormalized}.}
The RPA {and rPT2 methods} accurately model the vdW interactions
but also suffer from higher computational and basis set demands\cite{ren2012resolution, ren2012random, ren2013renormalized}.
Therefore, several other computationally efficient approximate models for the vdW dispersion interactions have emerged in recent decades\cite{klimevs2012perspective, stohr2019theory}.

Given the generality of Kohn-Sham theory, some methods attempt to incorporate nonlocal correlation directly into the density functional.
The vdW-DF of Dion \emph{et al.} \cite{dion2004van}, is a prototypical example of this approach. Although it suffered from inaccuracy on some systems \cite{vydrov2009nonlocal, klimes2011van}, the techniques that it introduced inspired other functionals that include vdW terms such as vdW-DF2, VV10, rVV10, SCAN+rVV10, and others \cite{lee2010higher, vydrov2010nonlocal, sabatini2013nonlocal, peng2016versatile, chakraborty2020next}.
These functionals incorporate a nonlocal dispersion term that depends on the density at two spatially separated points, which allows them to model long-range interaction \cite{berland2015van, tran2019nonlocal}, while methodological advances have reduced the computational demand of this nonlocality \cite{RomanPerez2009efficient, sabatini2013nonlocal}.
However, these methods can accuracy issues on some systems and may have inconsistent scaling with system size.\cite{hermann2018electronic}

Other approaches to modeling vdW interactions use an additive term $E_{\mathrm{DFT}} + E_{\mathrm{vdW}}$, which is usually calculated separately from the DFT self-consistent-field (SCF) procedure. The simplest versions of these methods use a pairwise scheme, where each A--B fragment pair in the system (e.g. atom pairs) contribute additively to the total energy \cite{kim2016recent}.
Under second-order perturbation theory, the vdW dispersion energy in non-conductive systems is proportional to ${{R_{AB}}^{-6}}$ to leading order, where $R_{AB}$ is the inter-fragment distance \cite{wang1927gegenseitige, london1937general, dobson2006asymptotics, lebgue2010cohesive}.
Thus, many vdW energy methods follow the form
\begin{equation}
E_\text{vdW} = -\frac{1}{2} \sum_{A,B} {f_\text{damp}(R_{AB},R_{A}^0,R_{B}^0) \cdot \frac{C_{6,AB}}{ {R_{AB}}^6}}
\label{eqn:normal_R-6vdW}
\end{equation}
where ${C_{6,AB}}$ is the ${C_6}$ coefficient of the A--B pair; ${f_{\mathrm{damp}}}$ is a damping function used to resolve the numerical singularity when ${R_{AB}}$ approaches zero; ${R_{A}^0}$ and ${R_{B}^0}$ are the free-atom atomic radii of A and B. Different choices of the $C_6$ coefficients and the damping function have been proposed.

Wu and Yang developed an early implementation of this type of vdW term, using an optimized set of species-dependent $C_6$ coefficients \cite{wu2002empirical}.
The commonly-used DFT-D family of methods introduced by Grimme \emph{et al.} was initially similar, using optimized static $C_6$ coefficients \cite{grimme2004accurate}, but later versions have been extended to consider the influence of local environment, charge redistribution, and three-body effects on the $C_6$ coefficients \cite{grimme2010consistent, caldeweyher2017extension}.
These methods are widely applicable, with errors on $C_6$ coefficients below \qty{5}{\percent} \cite{caldeweyher2017extension}.
The method can also be adapted through the damping scheme, as in the DFT-D(BJ) method \cite{grimme2011effect} or the optimized damping DFT-D(op) by Gordon \emph{et al.} \cite{witte2017assessing} leading to applications in non-covalent clusters, isomerization, and equilibrium geometry prediction.

While the DFT-D type methods use primarily geometrical descriptors (e.g. coordination number) to incorporate the effect of an atom's local environment into the long range dispersion, accurate results can also be obtained by using the electron density directly.
The pairwise model of Tkatchenko and Scheffler (TS\_2009), investigated in this work, uses an effective atomic volume calculated from the electron density to scale free-atom polarizabilities and $C_6$ coefficients to account
for the environmental dependence \cite{tkatchenko2009accurate}.
This scheme is widely applicable to both molecular systems and extended periodic systems.
{Similar in some ways, the exchange-hole dipole method also calculates the $C_n$ coefficients from the local electronic structure, but it does so by calculating atomic exchange-hole multipole moments instead of by rescaling free-atom reference data\cite{johnson2006post}.
The method also includes higher order terms and uses specific density functionals and damping terms to avoid double counting at short range\cite{price2021requirements}.}

Despite the success of pairwise models, such approximations do not include screening effects from the intervening material or long range charge oscillations (Dobson type B and C terms) \cite{dobson2014beyond}.
Intermediate pairwise approaches can incorporate some macroscopic screening effects\cite{tao2017screened}.
These effects are included in many-body methods such as RPA {and rPT2}, and are also partially included in many-body dispersion (MBD) type methods that model atoms as a system of coupled fluctuating dipoles \cite{tkatchenko2012accurate, tkatchenko2013interatomic, ambrosetti2014long, hermann2020density, kim2020umbd}.
More recent nonlocal versions (MBD-NL)\cite{hermann2020density} incorporate ideas from nonlocal DFT functionals \cite{vydrov2010dispersion} to improve treatment of the polarizability.

Due to its balance of methodological simplicity, robustness, and transferability, the TS\_2009 method nevertheless remains widely used.
For example, it has proven to be particularly effective in modeling systems such as hybrid perovskites that include both organic and inorganic subsystems, since these subsystems are partially connected by vdW interactions.\cite{ferjani2022first, liu2018tunable, heine2023benchmark}.
Some of our groups' past work shows that the Perdew--Burke--Ernzerhof (PBE) semi-local functional \cite{perdew1996generalized} with the TS\_2009 dispersion energy (PBE+TS\_2009) can give promising geometries (deviations within about \qty{2}{\percent} of the experimental structures) for hybrid organic--inorganic perovskite (HOIP) systems \cite{liu2018tunable, jana2019direct, gao2019molecular, lu2020highly, jana2020organic, song2023structure, park2023thickness, bhattacharya2023spin}.
Given the availability of more recent free-atom reference values since the original TS\_2009 implementation\cite{fedorov2018quantum}, with particularly large changes for alkali elements, we here study the performance of the updated TS method on alkali dimers, \num{45} inorganic solids, and three halide perovskites.
The primary aim here is to document {and assess} the TS\_2018 and TS\_alkali methods for use in practice, in an understandable and accessible form.
{Comparisons with the PBE+D3(BJ) method\cite{grimme2010consistent, grimme2011effect}, the B86bPBE+XDM method\cite{oterodelaroza2020application, price2023xdm}} and the beyond-pairwise MBD-NL approach\cite{hermann2020density} {establish their relative accuracy and identify remaining method-dependent differences in structural predictions.}

\section{Theory and Computational Methods}

{The six vdW methods assessed in this work represent several different paradigms.
The three TS methods are pairwise and follow Eq.~\ref{eqn:normal_R-6vdW}, and the $C_{6,AB}$ coefficients are calculated based on the local electronic structure.
The DFT-D3 method moves beyond a strictly pairwise method by including both pairwise terms to 6th and 8th order and three body terms. However, in this method, the $C_n$ coefficients are calculated using only the local geometry, rather than the electronic structure.
The XDM method includes only pairwise terms, but includes interactions up to 10th order and calculates the $C_n$ coefficients based on the exchange-hole multipoles calculated from the local electronic structure.
Finally, the MBD-NL method is explicitly nonlocal and beyond-pairwise.}
{For all employed methods, this work focuses on their physical accuracy, not computational efficiency or performance optimization.
Each method is employed as implemented in the FHI-aims all-electron electronic structure code,\cite{blum2009ab} the libMBD library,\cite{hermann2023libmbd} and the simple-dftd3 library.\cite{ehlert2024simple}
In Appendix~\ref{sec:timings} (Table~\ref{tab:timings}), we show that the computer time spent on the dispersion corrections remains essentially negligible compared to the overall compute time for a 40-atom hybrid metal halide perovskite unit cell as used elsewhere in this work.}
{We outline the TS methods to elaborate on the changes made in the updated TS\_2018 and TS\_alkali methods in the following.}

\subsection{Tkatchenko-Scheffler Dispersion Method}

The dispersion energy of Tkatchenko and Scheffler \cite{tkatchenko2009accurate} is a pairwise inter-atom vdW model developed by taking into account the influence of the local chemical environment. 
The effective ${C_6}$ coefficient each atom is given by scaling the free-atom coefficient, $C_{6,AA}^{\mathrm{free}}$, by comparing the effective atom volume ($V_{\mathrm{eff}}$) determined from a Hirshfeld partition of electron density and its corresponding free-atom volume ($V_{\mathrm{free}}$) \cite{brinck1993polarizability, olasz2007use}.
Free-atom dipole polarizabilities ($\alpha_A^{\mathrm{free}}$)\cite{chu2004linear} are similarly scaled to give effective atomic polarizability,
\begin{gather}
    C_{6,AA}^{\mathrm{eff}} = C_{6,AA}^{\mathrm{free}} \cdot 
    \left( \frac{V_{\mathrm{eff}}}{V_{\mathrm{free}}} \right)^2
    \label{eq:C6_scaling} \\
    \alpha_A^{\mathrm{eff}} = \alpha_A^{\mathrm{free}} \cdot 
    \left( \frac{V_{\mathrm{eff}}}{V_{\mathrm{free}}} \right)
    \label{eq:alpha_scaling}
\end{gather}
The heteronuclear interaction parameters are then calculated from these scaled coefficients according to
\begin{equation}
{C_{6,AB}} = \frac{2{C_{6,AA}^{\mathrm{eff}}}{C_{6,BB}^{\mathrm{eff}}}}
{
\left(\frac{\alpha_{B}^{\mathrm{eff}}}{\alpha_{A}^{\mathrm{eff}}}\right) 
{C_{6,AA}^{\mathrm{eff}}}
+ \left(\frac{\alpha_{A}^{\mathrm{eff}}}{\alpha_{B}^{\mathrm{eff}}}\right)
{C_{6,BB}^{\mathrm{eff}}}
}.
\label{eqn:C6_AB}
\end{equation}
A Fermi-type function\cite{wu2002empirical} is used for the damping function in Eq.~\ref{eqn:normal_R-6vdW},
\begin{equation}
f_{\mathrm{damp}} \left( R_{AB}, R_{AB}^0\right)
= \frac{1}{
    1 + \exp\left[
        -d \cdot \left( \frac{R_{AB}}{s_R \cdot R_{AB}^{0}} - 1 \right)
    \right]
}
\label{eqn:damping}
\end{equation}
where $R^0_{AB} = R_{A}^0+R_{B}^0$ and $R_{A}^0$ and $R_{B}^0$ are the pretabulated vdW radii of the chemical elements corresponding to atoms A and B, respectively.
The parameters $d$ and $s_R$ determine the steepness and onset radius of the vdW term, with $d = 20\,a_0$ ($a_0$ denotes the Bohr radius) and $s_R$ is an XC functional-specific empirical coefficient, e.g., $s_R = \num{0.94}$ for PBE\cite{perdew1996generalized} and $s_R = \num{0.96}$ for PBE0\cite{adamo1999toward}.
In addition to the relationships outlined in Equations \ref{eqn:normal_R-6vdW}, \ref{eqn:C6_AB}, and \ref{eqn:damping}, the TS\_2009 method also uses reference values for the homonuclear free-atom ${C_6}$ coefficients $C_{6,AA}^{\mathrm{free}}$, the free-atom static polarizability $\alpha_A^{\mathrm{free}}$, and the free-atom vdW radius ${R^0_A}$.
For this method, the $C_{6,AA}^{\mathrm{free}}$ and $\alpha_A^{\mathrm{free}}$ values calculated using time-dependent DFT in Ref.~\citenum{chu2004linear} are used as reference, 
with data for additional elements tabulated in appendix~A of Ref.~\citenum{gobre2016efficient}.
The vdW radii are defined as the radius of an isosurface in the free-atom electron density, where the value of this density contour is defined according to that of the noble gas in the same row \cite{tkatchenko2009accurate}.
Details of the performance of the non-self-consistent TS\_2009 dispersion energy with a broad range of representative XC functionals (including those based on the local density approximation and the generalized gradient approximation, conventional and screened hybrid functionals, M06 semi-empirical functionals, and nonempirical meta-GGA functionals) can be found in Ref.~\citenum{marom2011dispersion}.

\subsection{Revised Formula of Tkatchenko-Scheffler Dispersion Method}

\begin{figure*}
\centering 
\includegraphics{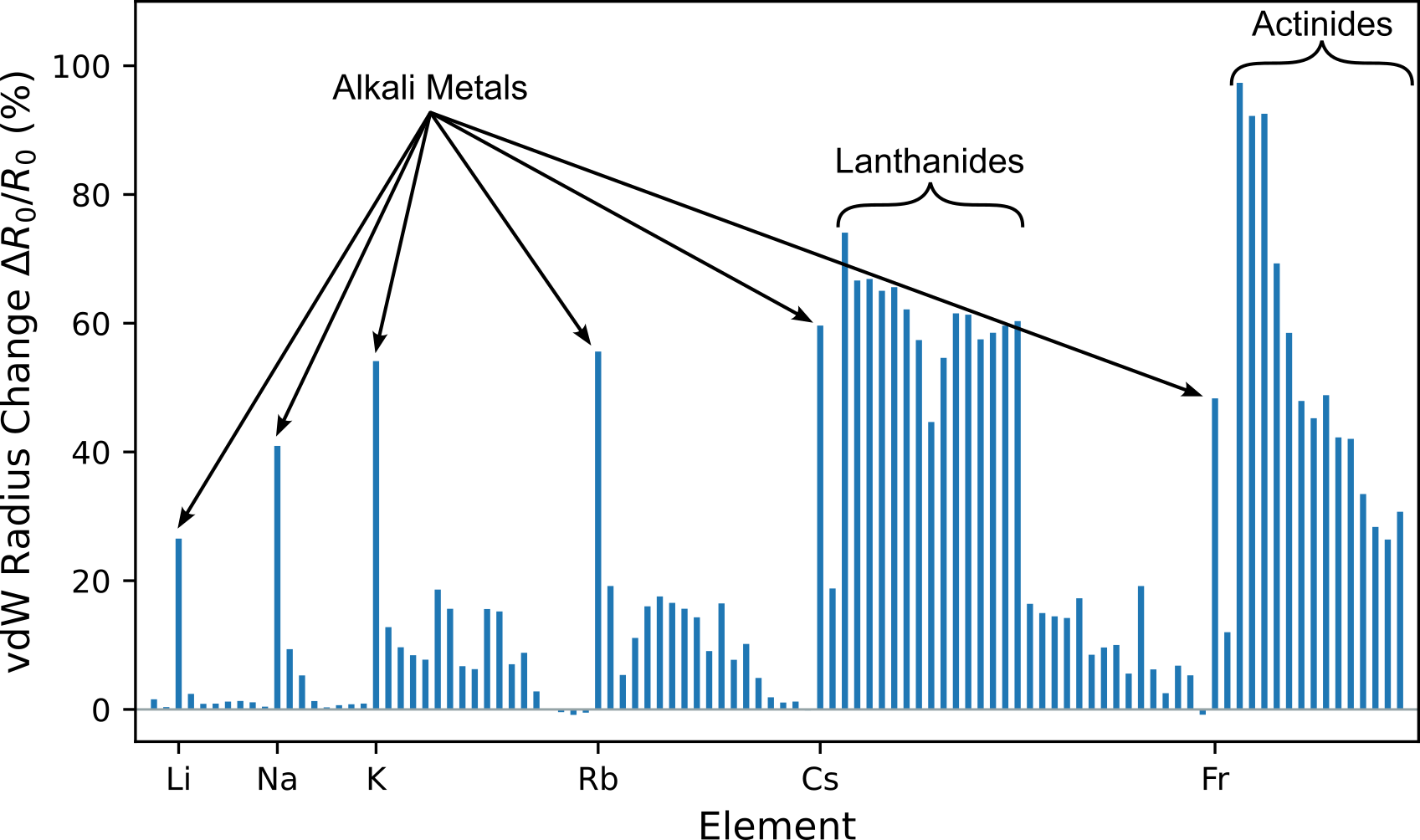}
\caption{Change of the free-atom vdW radii between TS\_2018 and TS\_2009 methods relative to the radius used in the TS\_2009 ($\Delta R_0 / R_0$). The most significant differences exist for alkali elements, lanthanide elements, and actinide elements.
All values are taken from Ref.~\citenum{gobre2016efficient}.}
\label{fig:Radii_summary}
\end{figure*}

Unlike the atomic polarizabilities and the related $C_6$ coefficients, which are observables and can be unambiguously determined in both theory and experiment, vdW radii are, in essence,  
empirically defined quantities. Past estimates arise from statistical analysis of systems in which the element in question exhibits characteristics similar to a closed-shell state (e.g.,as part of a closed-shell molecule).
In 2018, Fedorov \emph{et al.} revisited the definition of vdW free-atom radii more systematically from a fundamental perspective\cite{fedorov2018quantum}.
They derived a new simple quantum-mechanical scaling relation for the atomic vdW radius in terms of the dipole polarizability,
\begin{equation}
    R^0_{A} = C \cdot \left(\alpha_A^{\mathrm{eff}}\right)^{\frac{1}{7}}
    \label{eq:radius_scaling}
\end{equation}
where $\alpha_A^{\mathrm{eff}}$ is the effective atomic dipole polarizability defined in Eq.~\ref{eq:alpha_scaling} and $C$ is a constant.
Using numerical fitting of the vdW radii for noble gases \cite{bondi1964van, runeberg1998relativistic}, this constant was determined to be $C = \num{2.54}$ in atomic units\cite{fedorov2018quantum}.
A subsequent derivation has connected the proportionality constant $C$ to the fine structure constant.\cite{tkatchenko2021fine}
This gives a more physically meaningful definition of the vdW radius in terms of the atomic dipole polarizability, a quantity which can be determined accurately through experimental and theoretical methods.
This relationship can be used in calculating the atomic vdW radii necessary in Eq. \ref{eqn:normal_R-6vdW}, and it was validated with data on seventy-two elements from hydrogen to uranium. 
The deviation between the free-atom radii calculated using the electron density based definition of TS\_2009\cite{tkatchenko2009accurate} and the updated definition based on polarizability in Eq.~\ref{eq:radius_scaling} used in the TS\_2018\cite{fedorov2018quantum} method is summarized in Figure~\ref{fig:Radii_summary}.
While most main group and transition metals elements have similar vdW radii using both methods, the vdW radii of alkali elements, lanthanides, and actinides are all substantially larger when using the updated formula of TS\_2018.
We note that lanthanides and actinides are not included in the current benchmarks because the multireference nature of these elements would introduce a separate layer of complexity, distinct from the dispersion energy on its own.

Fedorov \emph{et al.} also revised the equilibrium vdW distances for heteronuclear dimers, $R^0_{AB}$ in Eq.~\ref{eqn:damping}\cite{fedorov2018quantum}.
Instead of directly summing the free-atom radii, the atomic polarizabilities are first averaged before calculating the vdW radius, so
\begin{equation}
    R_{AB}^{0, (\mathrm{TS\_2018})} = 2 \cdot 2.54 \cdot \left(
    \frac{{\alpha_A^{\mathrm{free}}} +  {\alpha_B^{\mathrm{free}}}}{2}
    \right) ^ \frac{1}{7}.
    \label{R0_AB_2018}
\end{equation}
The new TS\_2018 formula outperforms the TS\_2009  formula
\cite{fedorov2018quantum} with an average relative error, $\langle R.E. \rangle$, of 0.2$\%$ and an average absolute relative error, $\langle |R.E.| \rangle$, of 1$\%$ compared to experimental values of the equilibrium distances of the noble gas dimers \cite{tang2003van}.
In the TS\_2018 revision, the damping function was updated as well as the vdW radii, and these change were later implemented into the FHI-aims package\cite{blum2009ab,kim2022structural,abbott2026roadmap}, an all-electron electronic structure code that employs numerically tabulated atom-centered orbital basis sets.
An intermediate method (TS\_alkali) was also introduced, where the vdW radii of alkali metals were updated (as these were the light elements most strongly affected), while the form of the damping function and the vdW radii of other elements were not changed.\cite{kim2022structural}
Thus, in systems without alkali elements, TS\_alkali is equivalent to TS\_2009.

\section{Results and Discussion}

\subsection{Alkali Dimers}

Among all {non-f-block} elements, the alkali ones show the largest deviations in free-atom radii going from TS\_2009 to TS\_2018 (Figure~\ref{fig:Radii_summary}).\cite{kim2022structural} We therefore begin by examining the performance of the TS\_2018 method as implemented in the FHI-aims, with the simple case of alkali metal dimers, which are often used in the context of examining vdW methods.\cite{lima2005long,mitroy2007long,tao2010long,alves2010van}
We consider the potential energy curve of the dimers to compare the performance of the PBE+TS\_2018 and PBE+TS\_alkali with other methods. 
As a reference, the energy is calculated using the exact exchange and {renormalized second order perturbation theory (rPT2) for the correlation energy using the PBE-GGA KS orbitals (rPT2@PBE) \cite{ren2013renormalized}.}
{This method improves upon the random phase approximation and is discussed further in Appendix~\ref{sec:comput_details}.}
This reference is then used to evaluate the binding energy calculated using the the PBE functional without any additional vdW contribution \cite{perdew1996generalized}, the PBE functional combined with the pairwise PBE+TS\_2009 method\cite{tkatchenko2009accurate}, and with the PBE+TS\_2018 method\cite{fedorov2018quantum}.
In systems containing only a single type of alkali element, the TS\_alkali method is equivalent to the TS\_2018 method.
For comparison, {the energy was also calculated using the PBE+D3(BJ) method\cite{grimme2011effect}, the B86bPBE+XDM method\cite{price2023xdm}, and the nonlocal PBE+MBD-NL\cite{hermann2020density} method}.
The binding energy, $E_{\mathrm{bind}}$, as a function of the interatomic separation distance ($R$) is calculated as,
\begin{equation}
    E_{\mathrm{bind}}(R) = E_{\mathrm{tot},AA}(R) - 2 \cdot E_{\mathrm{tot},A}
    \label{eq:binding_energy}
\end{equation}
where $E_{\mathrm{tot},AA}(R)$ is the total energy of the dimer and $E_{\mathrm{tot},A}$ is the energy of an isolated atom. {A counterpoise correction is applied when calculating the rPT2@PBE binding energy;} further details are available in Appendix~\ref{sec:comput_details}.

\begin{figure*}
\centering 
\includegraphics{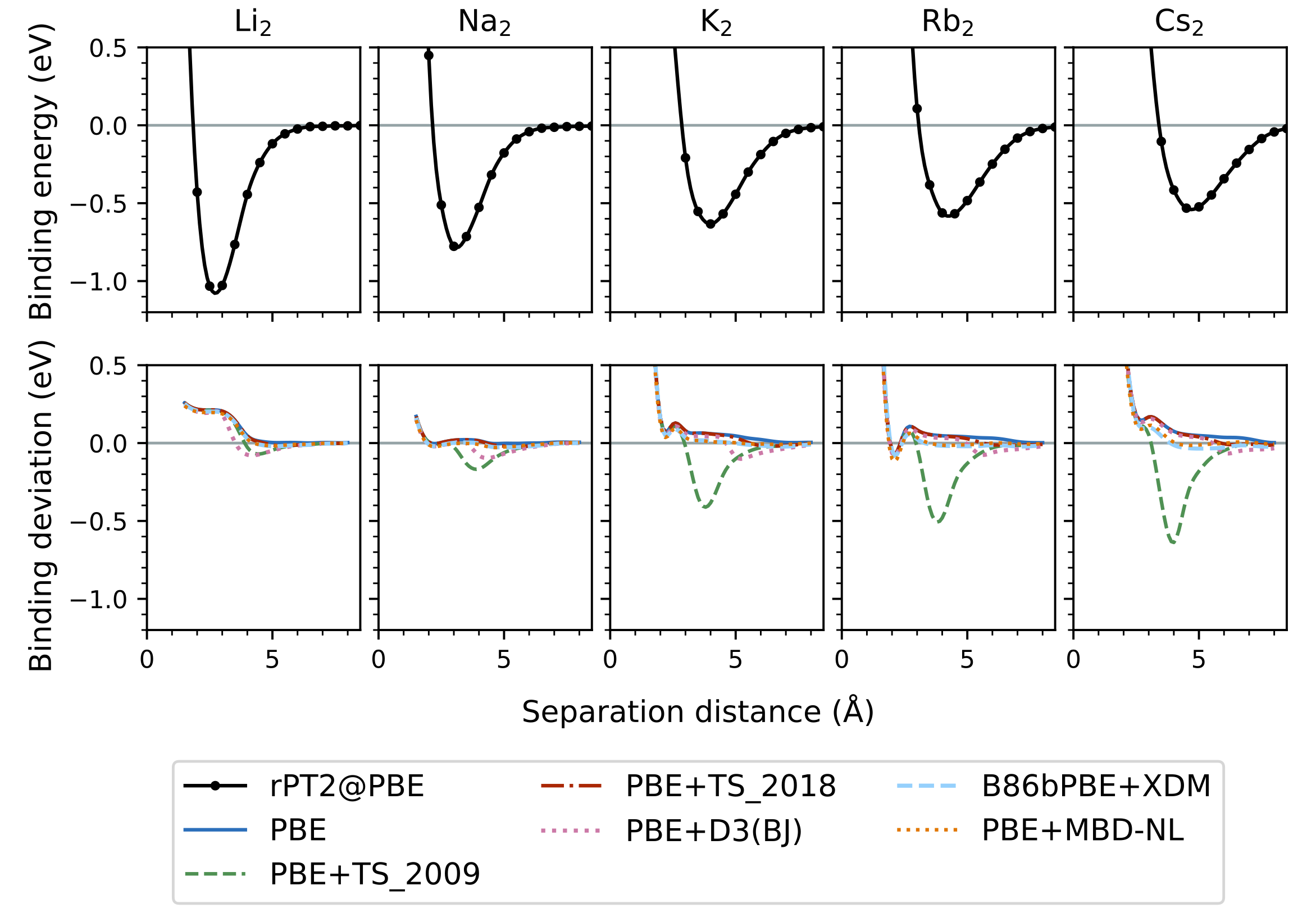}
\caption{{Top row: Reference binding energy curves of alkali metal dimers \ce{Li2}, \ce{Na2}, \ce{K2}, \ce{Rb2}, and \ce{Cs2}. The binding energies are calculated with rPT2@PBE using Eq.~\ref{eq:binding_energy}.
Bottom row: Deviation of the binding energies predicted by PBE, PBE+TS\_2009, PBE+TS\_2018, PBE+D3(BJ), B86bPBE+XDM, and PBE+MBD-NL from the rPT2@PBE reference for alkali metal dimers, defined as $E_{\mathrm{bind}}^{\mathrm{(vdW)}}(R) - E_{\mathrm{bind}}^{\mathrm{(rPT2)}}(R)$.}
Note that the TS\_alkali method is equivalent to the TS\_2018 method on these systems.}
\label{fig:dimer}
\end{figure*}

The {rPT2@PBE} binding energy curves of the alkali dimers \ce{Li2}, \ce{Na2}, \ce{K2}, \ce{Rb2} and \ce{Cs2} are shown in the top row of Figure~\ref{fig:dimer}.
All five cases show repulsion at the short distance and sizable attraction at the intermediate distance range.
The potential energy minimum shifts to longer distances when descending in the periodic table from \ce{Li2} to \ce{Cs2}, which is consistent with the increasing atomic radii.
In order to compare the performance of the tested vdW methods on these dimer systems, we show the deviations from the {rPT2@PBE} reference in Figure~\ref{fig:dimer}.
At short distances, all methods differ from the {rPT2@PBE} result, with larger differences at higher atomic number. This deviation is due to the PBE description of atomic repulsion at small distances and is not related to the vdW methods.
{All methods similarly underestimate the rPT2-predicted equilibrium binding energy of the \ce{Li2} dimer.}
For other dimers, at intermediate and long-distance ranges, {five methods (PBE, PBE+TS\_2018, PBE+TS\_alkali, B86bPBE+XDM, and PBE+MBD-NL) perform similarly}, with only small energy differences from the {rPT2@PBE} reference.
The good performance of uncorrected PBE on the dimer systems highlights that the bonding is not primarily vdW in character.
{The PBE+D3(BJ) method shows a slightly larger overbinding error for the \ce{Na2} and \ce{K2} dimers. However, it also develops a secondary feature at larger radii (Appendix~\ref{sec:binding_curves}, Fig.~\ref{fig:si-dimer-binding}). This behavior has also been observed in periodic alkali metals, where it is attributed to the coordination-number dependence of the $C_6$ coefficients\cite{kim2016failure, kim2020umbd}.}
On the other hand, Figure~\ref{fig:dimer} {and Figure~\ref{fig:si-dimer-binding}} show a shortcoming of the PBE+TS\_2009 method, as it significantly overestimates the binding energy in the intermediate distance range, {and in some dimers, the error is comparable to or larger than the reference binding energy}.
The over-binding becomes more pronounced from \ce{Li2} to \ce{Cs2}, and can be attributed to the underestimation of the free-atom radii for alkali metals in the TS\_2009\cite{tkatchenko2009accurate} vdW energy method.
This error in atomic radii and the corresponding over-binding is corrected in the TS\_2018\cite{fedorov2018quantum} and TS\_alkali schemes.

\subsection{Strongly bound solids}

\begin{figure*}
    \centering 
    \includegraphics{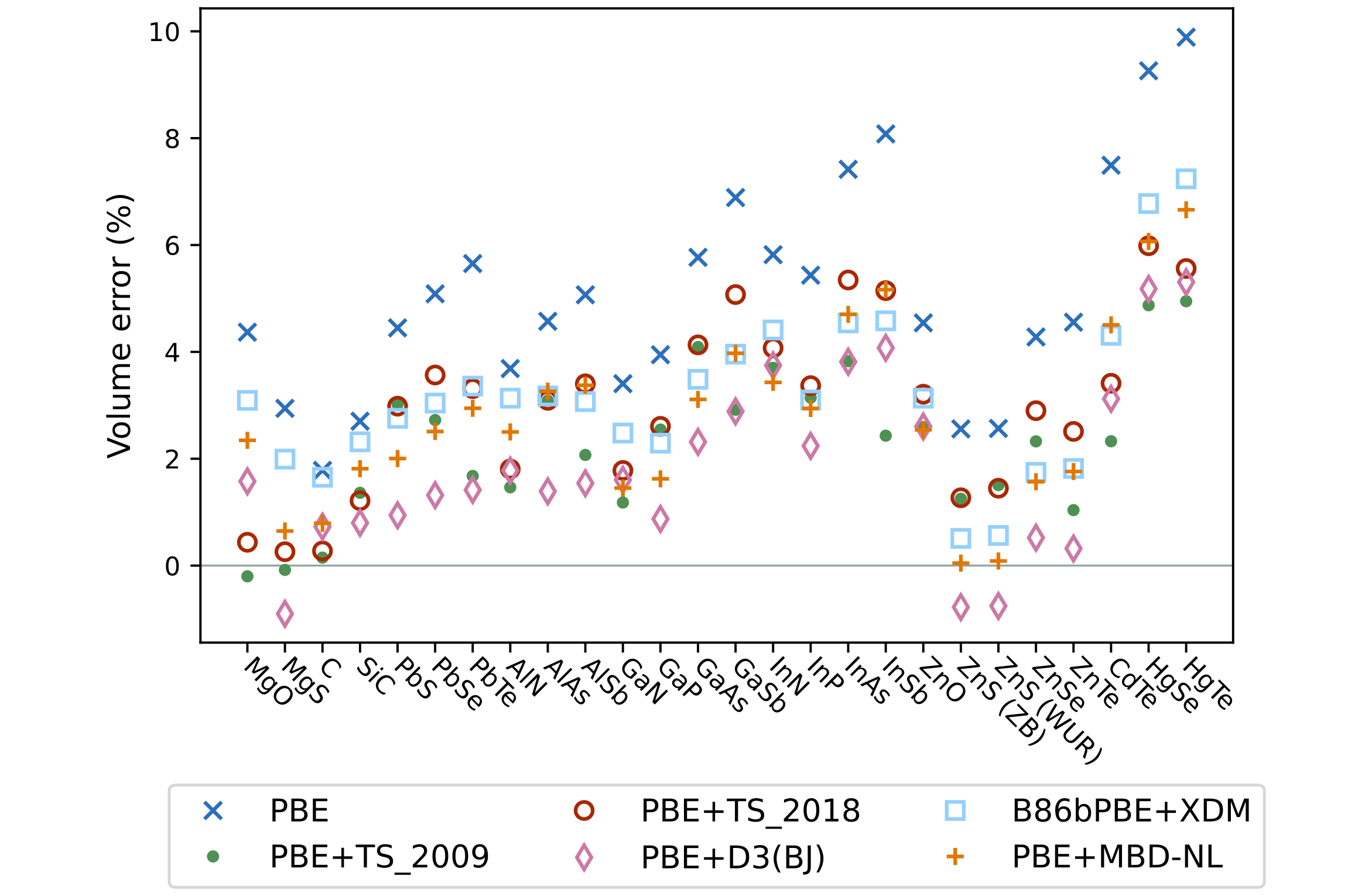}
    \caption{The relative unit cell volume difference between {the experimental geometry references and the relaxed geometries calculated using PBE, PBE+TS\_2009, PBE+TS\_2018, PBE+D3(BJ), B86bPBE+XDM, and PBE+MBD-NL for twenty-five non-alkali containing binary semiconductors and diamond.} For such systems, the TS\_alkali method is equivalent to the TS\_2009 method.
    For \ce{ZnS}, both zincblende (ZB) and wurtzite (WUR) structures are stable and are considered here. The reference volumes are discussed in Appendix~\ref{sec:solid_reference}.}
    \label{fig:semiconductors}
\end{figure*}

\begin{figure*}
    \centering 
    \includegraphics{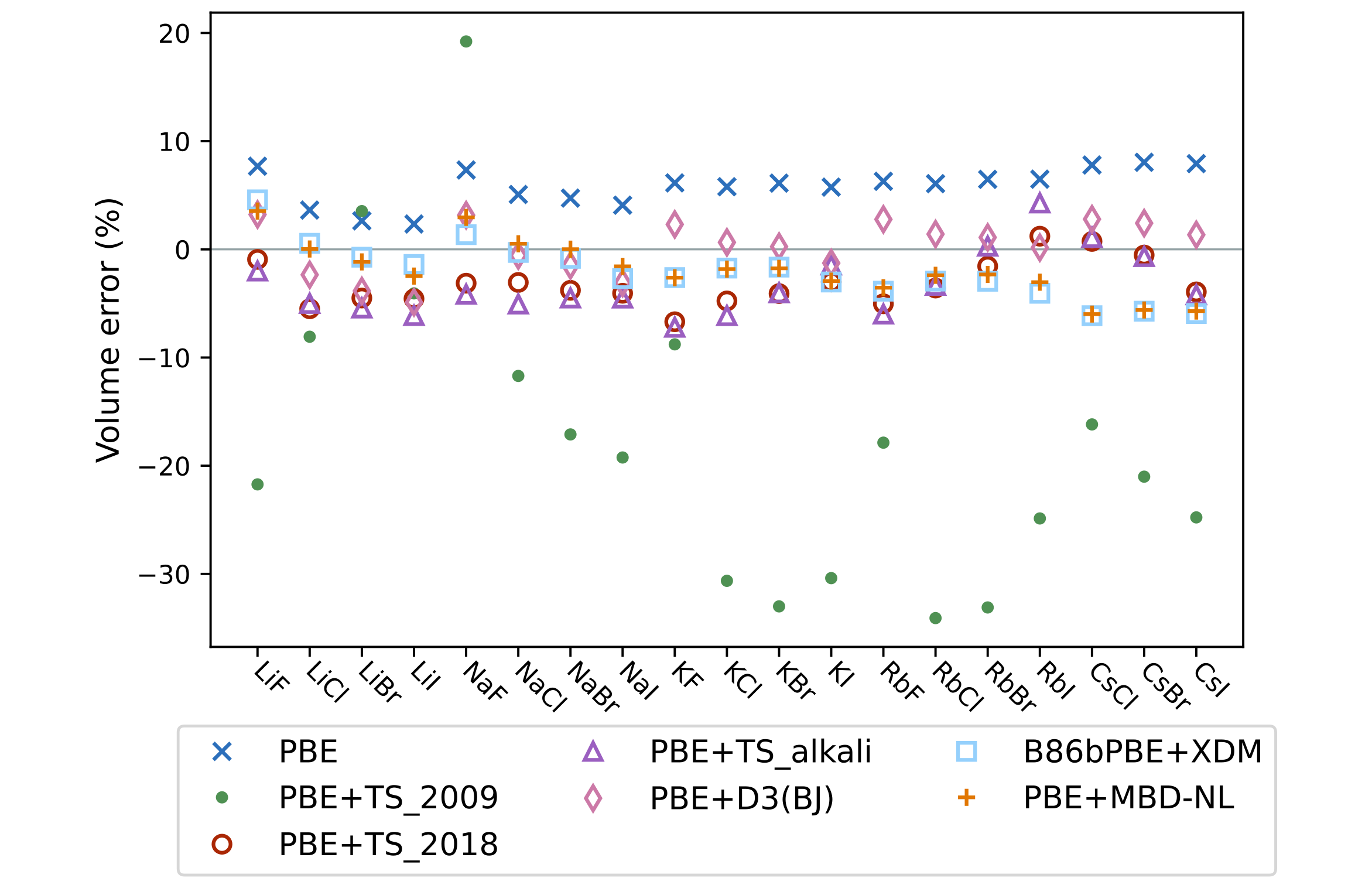}
    \caption{The relative unit cell volume difference between {the experimental geometry references and the relaxed geometries calculated using PBE, PBE+TS\_2009, PBE+TS\_2018, PBE+TS\_alkali, PBE+D3(BJ), B86bPBE+XDM, and PBE+MBD-NL for nineteen alkali halides.} The experimental references are discussed in Appendix~\ref{sec:solid_reference}.}
    \label{fig:alkali_halide}
\end{figure*}

To further examine the performance of the updated vdW energy calculations, we consider the effect of different vdW methods on the equilibrium structure of forty-five solid-state systems,  divided into non-alkali containing (Figure~\ref{fig:semiconductors}) and alkali-containing materials (Figure~\ref{fig:alkali_halide}).
For these systems, experimental structures are available, and structures corrected for zero-point motion are used as a reference (Appendix~\ref{sec:solid_reference}).
 
We perform geometry optimization using different vdW methods and use the relative difference in unit cell volume to assess the overall accuracy of the predicted structures (See Appendix~\ref{sec:comput_details} for computational details).
In the non-alkali containing materials of Figure~\ref{fig:semiconductors},
all vdW methods probed here yield improved results over the plain PBE functional.
For example, the uncorrected PBE functional makes an error of \qty{1.8}{\percent} in the volume of carbon in the diamond structure, {whereas the B86bPBE+XDM method makes a slightly smaller error of \qty{1.7}{\percent} and all other tested vdW methods reduce the error to $< \qty{1}{\percent}$}.

In addition to diamond, we examine the performance of different vdW methods on twenty-five binary semiconductors that do not contain alkali elements.
The unit cell volumes {calculated using the PBE+TS\_2009 (which is equivalent to TS\_alkali when alkali atoms are not present), PBE+TS\_2018, PBE+D3(BJ), B86bPBE+XDM, and PBE+MBD-NL methods are within \qty{7.5}{\percent} of the reference values for all systems} (Fig.~\ref{fig:semiconductors}).
The uncorrected PBE functional shows a consistently larger overestimation of the lattice volume.
{In contrast, all six vdw methods yield comparable results across the materials, with average absolute relative errors below \qty{3.5}{\percent} for each method (Table~\ref{tab:overall_relative_error}).}

We also calculate the structure of binary ionic compounds between alkali elements (\ce{Li}, \ce{Na}, \ce{K}, \ce{Rb} and \ce{Cs}) and halide elements (\ce{F}, \ce{Cl}, \ce{Br} and \ce{I}).
Due to the presence of alkali metal atoms, there are larger differences between the different calculation methods.
The performance of different vdW methods on these nineteen test cases is shown in Figure~\ref{fig:alkali_halide}.
The experimental data needed to calculate a correction for zero-point expansion is not available for \ce{CsF}, therefore a comparison is not included for this material.
The uncorrected PBE functional again overestimates the unit cell volume of each material.
In contrast, PBE+TS\_2009 systematically underestimates the unit cell volume on all alkali halides except for \ce{LiBr} and \ce{NaF}.
Consistent with the trend in the alkali dimers (Figure~\ref{fig:dimer}), the underestimation is most pronounced for the compounds containing \ce{K}, \ce{Rb}, and \ce{Cs}, where the error can exceed \qty{30}{\percent}.
This problem is largely corrected by PBE+TS\_2018 and TS\_alkali, which both provide {comparable performance to that of PBE+D3(BJ), B86bPBE+XDM, and PBE+MBD-NL}.
For highly-symmetric solids like the ones studied here, many-body contributions to dispersion interactions are essentially negligible because the electric field response is localized due to symmetry reasons. Hence, the similar performance of pairwise and MBD methods is expected.

The performance of {all tested vdW methods} in predicting the unit cell volume of simple solids is summarized in Table~\ref{tab:overall_relative_error}, which lists the relative error ($\langle R.E. \rangle$) and absolute relative error ($\langle |R.E.| \rangle$) of the cell volume averaged over materials with and without alkali halide atoms.
Unlike for the alkali dimers, in solid state systems, the systematic overestimation of the unit cell volume by the uncorrected PBE functional becomes apparent, with an average error over \qty{5}{\percent}.
Due to the large underestimation of free atom vdW radii of alkali elements, the PBE+TS\_2009 method does not perform well for alkali halide compounds, with an average absolute error in the unit cell volume of \qty{20}{\percent}.
The updated PBE+TS\_2018 method corrects the atomic vdW radii and thus largely eliminates errors stemming from the alkali elements, allowing the method to reach a similar accuracy as the other vdW methods.
This improvement carries over to the PBE+TS\_alkali method, which performs well on systems containing alkali atoms while otherwise maintaining exact compatibility with PBE+TS\_2009 results.
{The PBE+TS\_2018 and PBE+TS\_alkali methods, along with the PBE+D3(BJ), B86bPBE+XDM, and PBE+MBD-NL methods all provide reliable performance over the different simple binary compounds tested here.}

\begin{table}
\caption{The performance of {PBE, PBE+TS\_2009, PBE+TS\_2018, PBE+TS\_alkali, PBE+D3(BJ), B86bPBE+XDM, and PBE+MBD-NL} methods for geometry relaxations of alkali halides and semiconductor systems considered in this section. The average relative error ($\langle R.E. \rangle$) and the average absolute relative error ($\langle |R.E.| \rangle$) values show the overall deviation magnitude between predicted unit cell volumes from the experimental references.}
\label{tab:overall_relative_error}
\centering
\begin{ruledtabular}
\begin{tabular}{l d{3.2} d{2} d{2} d{2}}
& \multicolumn{2}{c}{Alkali Halides} & \multicolumn{2}{c}{Semiconductors} \\
\cmidrule(lr){2-3}
\cmidrule(lr){4-5}
{vdW method}
& \multicolumn{1}{c}{$\langle R.E. \rangle$} 
& \multicolumn{1}{c}{$\langle |R.E.| \rangle$}
& \multicolumn{1}{c}{$\langle R.E. \rangle$}
& \multicolumn{1}{c}{$\langle |R.E.| \rangle$} \\
\midrule
PBE            &  5.81  & 5.81  & 5.08 & 5.08 \\
PBE+TS\_2009   & -17.57 & 19.97 & 2.30 & 2.32 \\
PBE+TS\_alkali & -3.38  & 3.97  & 2.30 & 2.32 \\
PBE+TS\_2018   & -3.20  & 3.40  & 3.00 & 3.00 \\
 PBE+D3(BJ)     &  0.21  &  2.06  &  1.83 &  2.02 \\
 B86bPBE+XDM    & -2.11  & 2.79  & 3.17 & 3.17 \\
PBE+MBD-NL    & -1.89  & 2.63  & 2.76 & 2.76 \\
\end{tabular}
\end{ruledtabular}
\end{table}

\subsection{Cs-containing 3D Metal Halide Perovskites: \ce{CsPbCl3}, \ce{CsPbBr3} and \ce{CsPbI3}}

{We study a series of Cs-containing lead-halide perovskites in order to evaluate the impact of the vdW method on the predicted structure of these more complex materials. 
In particular, we demonstrate the change due to the improved description of alkali elements, especially \ce{Cs}, in the TS\_2018 and TS\_alkali methods compared to the original TS\_2009 method.}
We consider the three-dimensional perovskites \ce{CsPbCl3}, \ce{CsPbBr3}, and \ce{CsPbI3}.
Accurate prediction of the geometry of these systems while including the vdW interaction is critical for modeling hybrid perovskites where alkali cations and organic components coexist.\cite{kim2022structural, li2016stabilizing}
{Additionally, the choice of vdW method can influence the accuracy of predicted phase transitions in finite-temperature simulations.\cite{knysh2026bridging}}

\begin{figure*}
\centering
\includegraphics{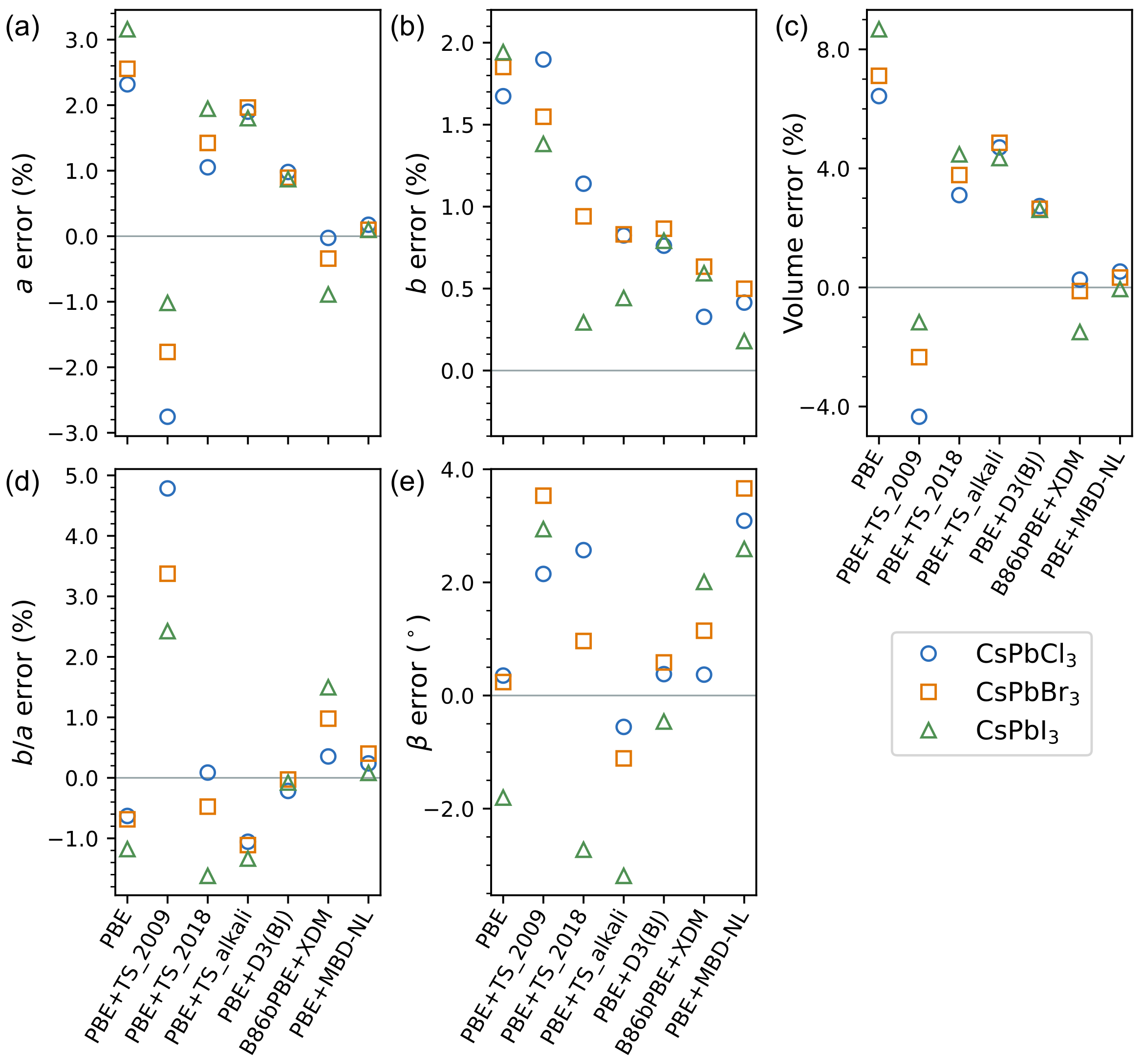}
\caption{Errors in the (a) $a$ lattice constant (expressed in the 40-atom supercell, where it is equal to the $c$ lattice constant in the $Pnma$ space group), (b) the $b$ lattice constant, (c) the unit cell volume, (d) the $b/a$ ratio, and (e) the $\beta$ angle of \ce{CsPbCl3}, \ce{CsPbBr3}, and \ce{CsPbI3} unit cells calculated using different vdW methods. Errors are relative to experimentally determined crystal structures.\cite{linaburg2017cs1, stoumpos2013crystal, straus2020understanding, muscarella2023which} In \ce{CsPbCl3}, the PBE+TS\_2009 functional leads to an additional symmetry breaking of the structure to $P2_1/m$ and therefore unequal $a$ and $c$ values; in Figure (a), the deviation of the $a$ value is shown for PBE+TS\_2009.}
\label{fig:perov_lat_error}
\end{figure*}

Starting with the experimental crystal structures (orthorhombic \ce{CsPbCl3}\cite{linaburg2017cs1, muscarella2023which} \ce{CsPbBr3}\cite{stoumpos2013crystal, muscarella2023which} and \ce{CsPbI3} unit cells\cite{straus2020understanding}), we relaxed the atomic positions and lattice parameters using different vdW methods (See Appendix~\ref{sec:comput_details}).
The primitive unit cell contains \num{20} atoms with the principle axis along the $b$ direction.
To {better capture potential symmetry breaking}, all calculations were performed in a \num{40} atom \numproduct[parse-numbers = false]{\sqrt{2}x1x\sqrt{2}} supercell.
In the supercell, an orthorhombic distortion results in a change in the $\beta$ angle between the supercell $a$ and $c$ axes.
Both the computationally relaxed lattice parameters and the experimental parameters are presented in Table~\ref{tab:perovskite_lattice_parameters}. We note that we here compare directly to cryogenic-temperature data ($T = \qty{90}{\K}$ or \qty{100}{\K}), rather than to a value in which the remaining impact of finite-temperature  and zero-point vibrational lattice expansion is corrected out. For the halide perovskites, the lattice expansion that remains within the systems at the chosen experimental reference temperature values could be affected by anharmonicities of the potential energy surface that are difficult to estimate precisely. In the quasiharmonic approximation and for the related halide perovskite methylammonium lead iodide, we have previously shown that the remaining volume contraction between $T=\qty{100}{\K}$ and the equilibrium value at the Born-Oppenheimer surface amounts to around \qty{1.5}{\percent}.\cite{heine2023benchmark}
For all methods except PBE+TS\_2009, the relaxed structure was in the same space group ($Pnma$) as the experimental structure, but the the lattice vectors ($a$, $b$, and $c$) and the $\beta$ lattice angle differ.
The PBE+TS\_2009 method incorrectly lowers the symmetry of \ce{CsPbCl3} to $P2_1/m$.
The calculations of the $b$ lattice vector show deviations of $<\qty{2}{\percent}$ for all vdW methods tested (Fig.~\ref{fig:perov_lat_error}b).
The calculated values of the $a$ and $c$ lattice vectors have slightly larger deviations (Fig.~\ref{fig:perov_lat_error}a), with the largest with PBE+TS\_2009 method underestimating the $a$ axis length of \ce{CsPbCl3} by \qty{2.8}{\percent} (because of the additional symmetry breaking in the optimized PBE+TS\_2009 structure, the $c$ axis length deviates by \qty{3.3}{\percent}).
{All other vdW methods make smaller errors, with the error in the predicted lattice constant below \qty{2}{\percent} in all other cases.}

To better understand these deviations from the experimental crystal structure, we show the errors in the cell volumes ($V$), the degree of tetragonality ($b/a$), and the $\beta$ angle in Figures~\ref{fig:perov_lat_error}(c-e) and Table~\ref{tab:volume_and_anisotropy}.
From these results, we see that the PBE functional without a vdW term consistently overestimates the cell volume (\qtylist[retain-explicit-plus]{+6.43; +7.11; +8.67}{\percent} for the three perovskite systems considered).
All vdW methods show a better agreement with the experimental volumes, with the PBE+MBD-NL method performing best for the selected experimental structures at $T\approx$ \qtyrange[range-phrase = --]{90}{100}{\K}.
Like in the binary halides, the PBE+TS\_2009 method consistently underestimates the perovskite cell volume, with an error of \qty{-4.3}{\percent} for the \ce{CsPbCl3} structure and closer agreement for the other structures. Since the computational values are calculated at the Born-Oppenheimer surfaces (no zero-point or finite-temperature lattice expansion included), the magnitude of the volume error made by the PBE+TS\_2009 functional might be smaller by another $\approx\qty{1.5}{\percent}$, as mentioned above. Interestingly, this would bring the PBE+TS\_2009 predicted volumes into somewhat better agreement with experiment compared to the bars shown in Figure~\ref{fig:perov_lat_error}c, especially for \ce{CsPbBr3} and \ce{CsPbI3}, rivalling the PBE+MBD-NL method (which would acquire a corresponding small overestimation). However, the internal structure of the cesium lead halide perovskites would nevertheless be unphysically distorted by PBE+TS\_2009, as we have observed previously for \ce{CsPbBr3} in Ref.~\citenum{kim2022structural}.
{The other vdW methods either overestimate the cell volumes, in the case of PBE+TS\_2018, PBE+TS\_alkali, and PBE+D3(BJ), or predict essentially correct volumes, in the case of B86bPBE+XDM and PBE+MBD-NL.}

The anisotropy of the perovskite cell is characterized by the $b/a$ ratio, which reflects the degree of tetragonality, and the $\beta$ angle, which is linked to the orthorhombic distortion (Fig.~\ref{fig:perov_lat_error}).
The PBE+TS\_2009 method exaggerates the cell anisotropy in all three systems, with both the $b/a$ ratio and the $\beta$ angle overestimated in all cases.
{This method makes an error of over \ang{3.5} on the $\beta$ angle in \ce{CsPbBr3} and an error of \qty{4.8}{\percent} on the $b/a$ ratio in \ce{CsPbCl3}.}
Other methods (including the uncorrected PBE) better capture the tetragonality, with {PBE+D3(BJ) and} PBE+MBD-NL performing particularly well.
{On the other hand, PBE+MBD-NL overestimates the orthorhombic distortion, making an error of \ang{3.7} on the $\beta$ angle of \ce{CsPbBr3} and similar errors for other perovskites.}
{For other methods, the results are more mixed, as several capture the orthorhombic distortion correctly for \ce{CsPbCl3} and \ce{CsPbBr3}, but make larger errors for \ce{CsPbI3} in its metastable perovskite phase.}

Overall, among the methods including vdW terms, PBE+TS\_2018 and PBE+TS\_alkali both give better predictions for halide perovskites when compared to the exaggerated anisotropy given by PBE+TS\_2009.
For PBE+MBD-NL, the result is mixed but overall favorable: the method correctly accounts for unit cell parameters of all three tested cesium lead halides with the exception of the $\beta$ angle.
{The PBE+D3(BJ) method performs consistently well in terms of cell shape, while it somewhat overestimates the volume, and the B86bPBE+XDM method also performs well in terms of both the cell shape and cell volume.}
Interestingly, the large thermal expansion associated with the halide perovskites leads to a separate conundrum, which is that their room-temperature lattice parameters are significantly higher\cite{muscarella2023which,straus2020understanding,heine2023benchmark} than the Born-Oppenheimer surface ones. Thus, the PBE+TS\_2018 and PBE+TS\_alkali approaches (which correctly reflect the local lattice symmetry of the cesium lead halides, compared to the artificial symmetry breaking in PBE+TS\_2009) also reflect the effective lattice parameter better at room temperature, though not for the right microscopic reasons.

\section{Conclusion}

{In this work, we benchmark two updated variants of the widely used Tkatchenko-Scheffler pairwise dispersion energy expression, paired with the even more widely used DFT-PBE\cite{perdew1996generalized} generalized gradient functional.
In particular, we show that these updated expressions (dubbed and TS\_alkali and TS\_2018, respectively) both correct predicted overbinding associated with alkali atoms, compared to the original approach (here called TS\_2009) published in 2009.\cite{tkatchenko2009accurate}
The narrower correction, TS\_alkali, retains the full parameterization of TS\_2009 except for the free-atom van der Waals radii associated with the alkali elements (these new radii are derived from an updated parameterization by Fedorov \emph{et al.} in 2018 \cite{fedorov2018quantum}).
Thus, the TS\_alkali approach remains exactly consistent with the original TS\_2009 approach, enabling full compatibility with the wide range of existing results in the literature that are based upon TS\_2009. In contrast, the TS\_2018 approach incorporates the complete updated parameterization of Fedorov \emph{et al.}, here benchmarked for a set of covalently and ionically bonded solids, including alkali-containing ones.
The TS\_2018 approach is shown to provide accuracy that is consistent with the original TS\_2009 approach for this set of materials, while also resolving any issues related to the alkali elements.
These benchmarks are important for placing any future application of the updated methods on sound footing; we demonstrate this specifically for the structural properties of  several widely studied cesium-based halide perovskites.}

{We furthermore compare the accuracy of the revised TS\_alkali and TS\_2018 approaches with that of several newer dispersion energy expressions, namely the D3(BJ) approach\cite{grimme2011effect}, the exchange-hole dipole moment (XDM)\cite{price2023xdm} method and the non-local many-body dispersion (MBD-NL) approach\cite{hermann2020density}, each paired with generalized-gradient functionals (PBE for D3(BJ) and MBD-NL, B86bPBE for XDM) with which the respective dispersion energy expressions are frequently employed.
We find that, for the inorganic solids considered here, all three provide overall similar, somewhat improved predictive accuracy over the updated Tkatchenko-Scheffler type approaches.
For halide perovskites in particular, it is clear that dispersion energy terms should be included in any case, i.e., even for the fully inorganic ones, in addition to the baseline GGA functional.}

{We deliberately focused our study on dispersion energy expressions paired with the computationally relatively efficient GGA functionals that still form a sweet spot for many practical production materials simulations carried out today.
In terms of accuracy improvements, it is well known that more sophisticated base density functionals, such as hybrid functionals or others that do not account for long-range dispersion  contributions on their own, are likely to improve upon some of the remaining small physical accuracy limits exhibited by the GGA based base functionals.
This is well documented, e.g., in the communities focused on molecular structure\cite{rossi2014validation,schubert2015exploring} and molecular crystal structure.\cite{firaha2023predicting,price2023xdm} 
Considering these more expensive prescriptions was not a focus of this work but does remain an important future direction, especially as large-scale efficiency\cite{kokott2024efficient} and applicability of higher-level functionals continue to evolve.}

\begin{acknowledgements}
The authors acknowledge Xinguo Ren for helpful discussion.
This work was funded by the U.S. National Science Foundation under awards No. DMR-2323803 and DMR-2323804.
Research Computing at the University of North Carolina at Chapel Hill is acknowledged for providing computational resources.
\end{acknowledgements}

\section{Data availability}

{All calculation data, including input and output files, processed data, and benchmark data gathered from references 
are openly available on Zenodo at
http://doi.org/10.5281/zenodo.23354573,
reference number~\citenum{bench_data}.}
Implementations of the TS\_2018 and TS\_alkali methods in the open-source libMBD package\cite{hermann2023libmbd} are forthcoming.

\appendix

\section{Computational details}
\label{sec:comput_details}

\subsection{Alkali dimers}

\begin{figure}
\centering
\includegraphics{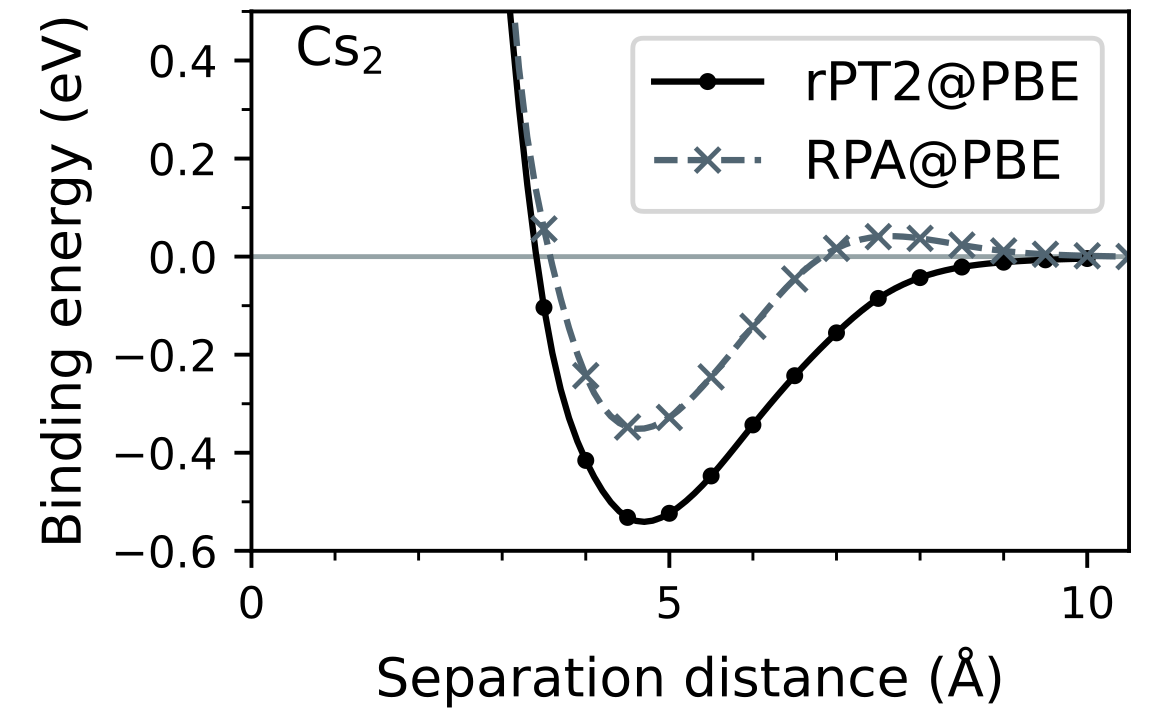}
\caption{{Comparison of binding energies of the \ce{Cs2} dimer calculated using the RPA@PBE and rPT2@PBE methods. Both calculations use the modified Tier~3 basis and the counterpoise correction discussed below.}}
\label{fig:si-rpa_compare}
\end{figure}

The {rPT2@PBE} reference energy includes the exact exchange and the {renormalized second order perturbation theory approximation for the correlation energy using the PBE Kohn-Sham orbitals\cite{ren2013renormalized}.}
{The rPT2 approach includes second order screened exchange (SOSEX) \cite{paier2010hybrid, *paier2010erratum, gruneis2009making} and renormalized single excitation (rSE) \cite{ren2011beyond} terms, in addition to the RPA contribution to the correlation
\cite{furche2001molecular, galano2006new, ren2012resolution, ren2012random} (i.e. $E_c^\mathrm{rPT2} = E_c^\mathrm{RPA} +  E_c^\mathrm{rSE} + E_c^\mathrm{SOSEX}$).}
{The rPT2 method resolves a known shortcoming where the RPA alone predicts an unphysical repulsion in the dimer singlet state at large but finite separation (Fig.~\ref{fig:si-rpa_compare})\cite{fuchs2005describing}.}

The binding energy $E_{\mathrm{bind}}$ as a function of the interatomic separation distance $R$ is calculated as
\begin{equation}
    E_{\mathrm{bind}}(R) = E_{\mathrm{tot},AA}(R) - 2 \cdot E_{\mathrm{tot},A} ,
    \label{eq:si-binding_energy}
\end{equation}
where $E_{\mathrm{tot},AA}(R)$ is the total energy of the alkali dimer and $E_{\mathrm{tot},A}$ is the energy of an isolated atom.
The {rPT2@PBE} reference binding energy was calculated using the FHI-aims code\cite{blum2009ab, ren2012resolution, ren2013renormalized, abbott2026roadmap} including spin polarization with the \texttt{really\_tight} numerical defaults.
The integration grids were modified by increasing both the number of radial grid point shells around each atom (by increasing the \texttt{radial\_multiplier} from 2 to 8) and the density of angular grid points on the outer shells (by increasing the maximum number of angular grid points in the outermost shell to \num{974}).\cite{delley1996high}
The onset of the confining potential for numerical atomic orbital (NAO) basis functions ($r_\mathrm{onset}$ in Eq.~9 of Ref.~\citenum{blum2009ab}) was expanded to \qty{12}{\angstrom}.

The basis set superposition error (BSSE) occurs when using atom-centered basis sets if the description of the electronic structure around an atom is improved by the presence of basis functions from neighboring centers.
It is often more significant for higher levels of electronic structure theory, such as the {rPT2@PBE} method used here, due to their slower convergence with basis set size.
The counterpoise correction is a practical scheme to largely eliminate BSSE for calculating binding energies.\cite{gutowski1986basis}
A counterpoise correction is applied such that the basis functions of the other atom in the dimer at a distance ${R}$ are also included when calculating the total energy of the individual atom.
Hence, the $E_{\mathrm{tot},A}$ term in Eq.~\ref{eq:si-binding_energy} becomes a function of inter-atomic distance, $E_{\mathrm{tot},A}(R)$.

\begin{figure*}
\centering 
\includegraphics{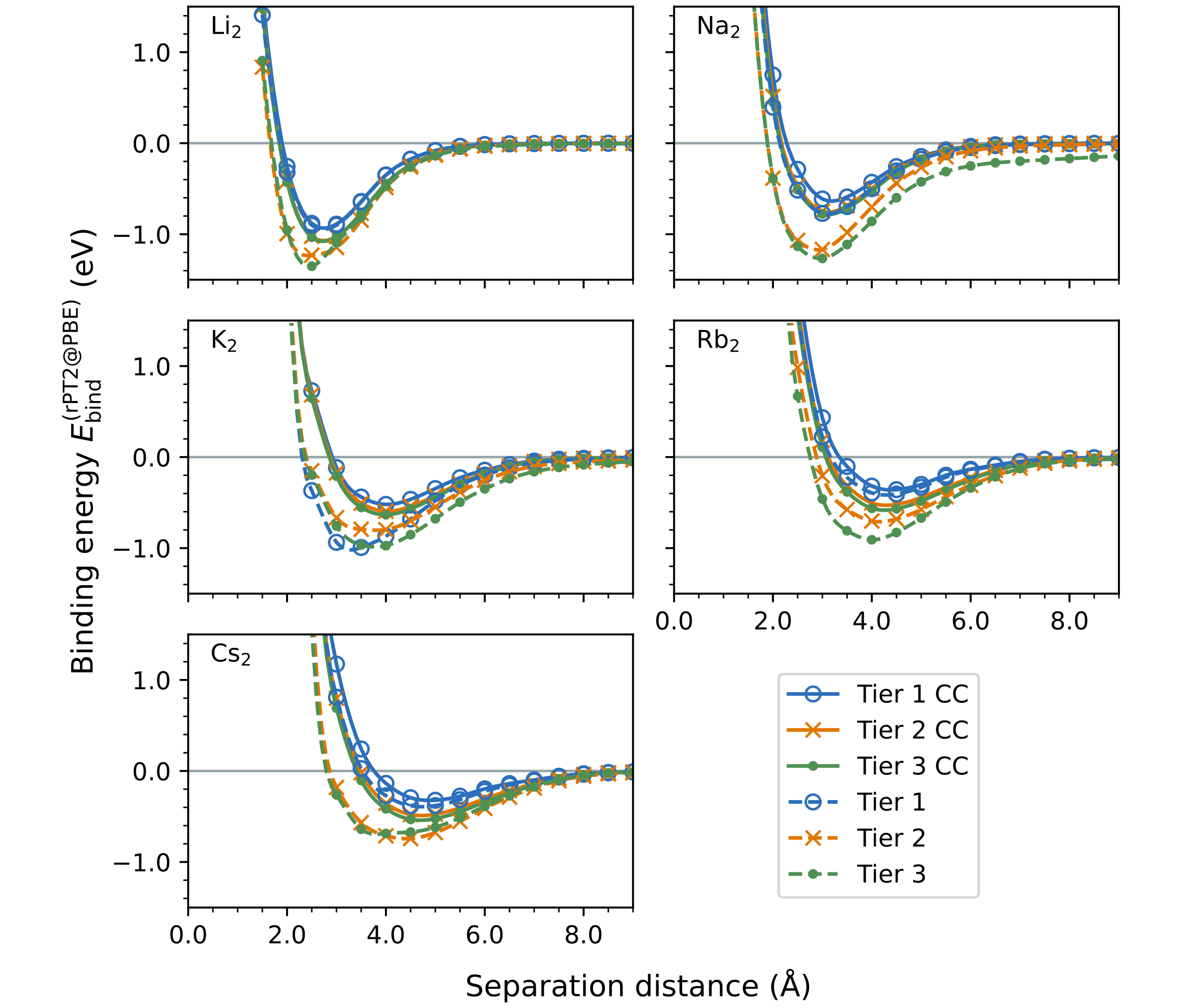}
\caption{Binding energy of alkali dimers calculated at the {rPT2@PBE} level using different basis sets, with and without counterpoise correction (CC).}
\label{fig:si-bind-rpa}
\end{figure*}

\begin{figure*}
\centering 
\includegraphics{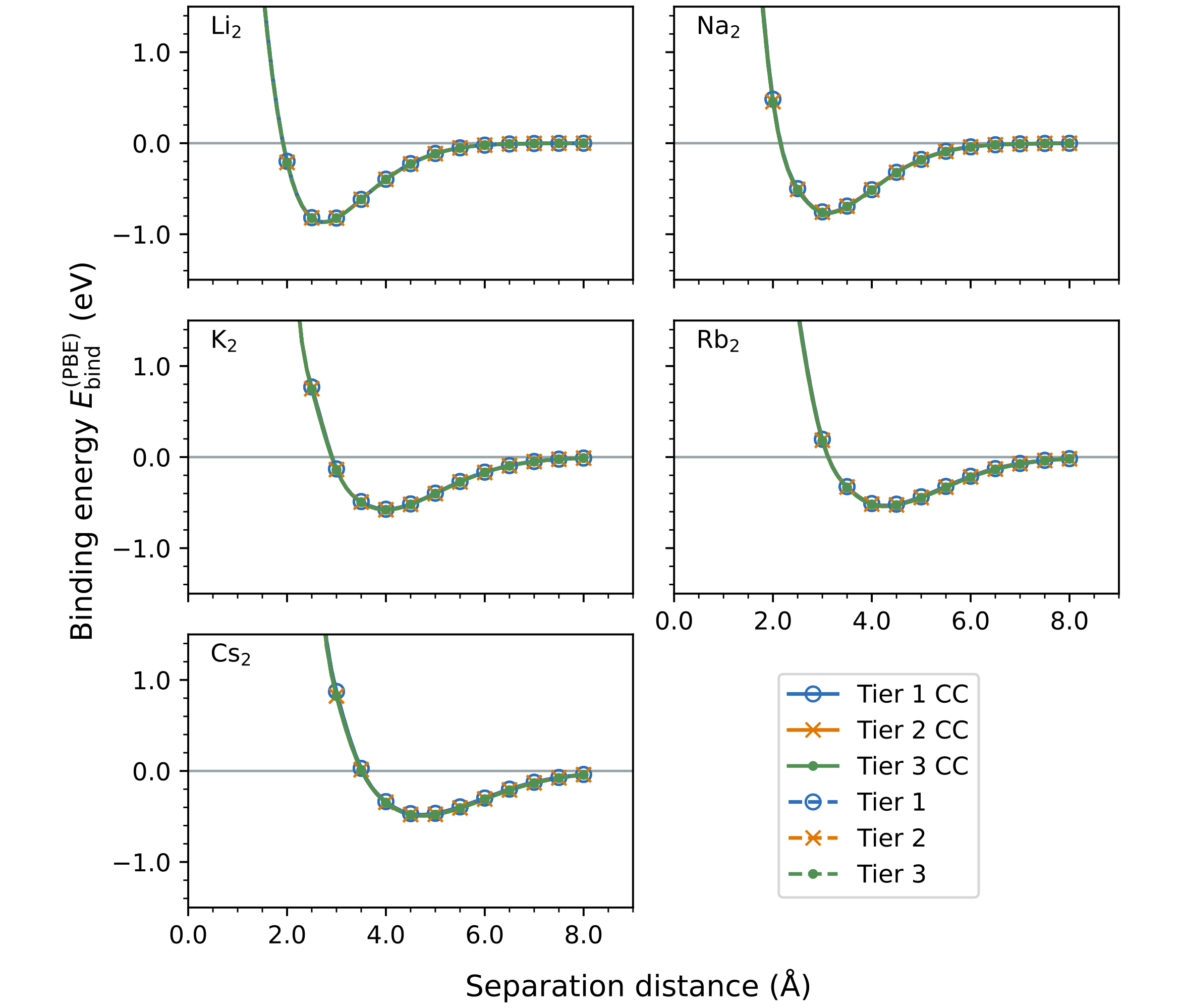}
\caption{Binding energy of alkali dimers calculated at the PBE level using different basis sets, with and without counterpoise correction (CC).}
\label{fig:si-bind-pbe}
\end{figure*}

As seen in Fig.~\ref{fig:si-bind-rpa}, the binding energy at the {rPT2@PBE} level without counterpoise correction is not converged even with a large ``Tier 3'' NAO basis.
However, when including the counterpoise correction, the results are well-converged.
Therefore we use a tier 3 basis and the counterpoise correction for the {rPT2@PBE} binding energies used in the main text.
In contrast, the binding energy calculated using PBE (Fig.~\ref{fig:si-bind-pbe}) is well-converged even with the tier 1 basis.
The counterpoise correction also has little influence on the binding energy.
The binding energies using PBE and methods including vdW terms were therefore calculated with the \texttt{tight} numerical defaults (which generally uses a Tier 1 or Tier 2 basis set) without the counterpoise correction. This numerical choice is widely used for reliable production calculations.

\subsection{Strongly bound solids}

All optimized geometries were calculated using the FHI-aims code without spin polarization using the \texttt{tight} numerical default basis set.
The Brillouin zone was sampled using a \numproduct{16x16x16} $\Gamma$-centered k-point grid.
Ground state volumes were calculated by fitting an equation of state following the procedure of Ref.~\citenum{lejaeghere2013error}.
For cubic systems, we optimize the atomic positions for 13 fixed volumes $V_0 \pm \qty{6}{\percent}$ around the experimental volume $V_0$ until forces were below \qty{5}{\meV/\angstrom}.
These points are then used to fit a Burch-Murnaghan equation of state.
The calculations and fitting procedure are then repeated for volumes around the $V_0$ fit in the first step.
For the alkali halides calculated with PBE+TS\_2009, the fitting procedure was performed thrice due to the large change in cell volume.
For the wurtzite structure, an approximate scheme is pursued. The experimental unit cell is first relaxed until effective forces on the lattice vectors are below \qty{5}{\meV/\angstrom}\cite{knuth2015strain}.
An equation of state is then fit as discussed above while keeping the shape of the unit cell fixed.

\subsection{Halide perovskites}

All calculations were performed in the conventional 40 atom unit cell.
Optimized geometries were calculated using the FHI-aims code with the ``tight'' numerical defaults and associated basis set.
However, due to the presence of heavy elements, a finer radial integration grid was used by increasing the \texttt{radial\_multiplier} from 2 (``tight'' default) to 4.
The Brillouin zone was sampled with a \numproduct{4x4x4} $\Gamma$-centered grid.
Atomic positions and lattice parameters were relaxed until forces were below \qty{1}{\meV/\angstrom}.

\clearpage
\onecolumngrid
\section{Dimer binding curves}
\label{sec:binding_curves}

\begin{figure*}[h]
\centering
\includegraphics{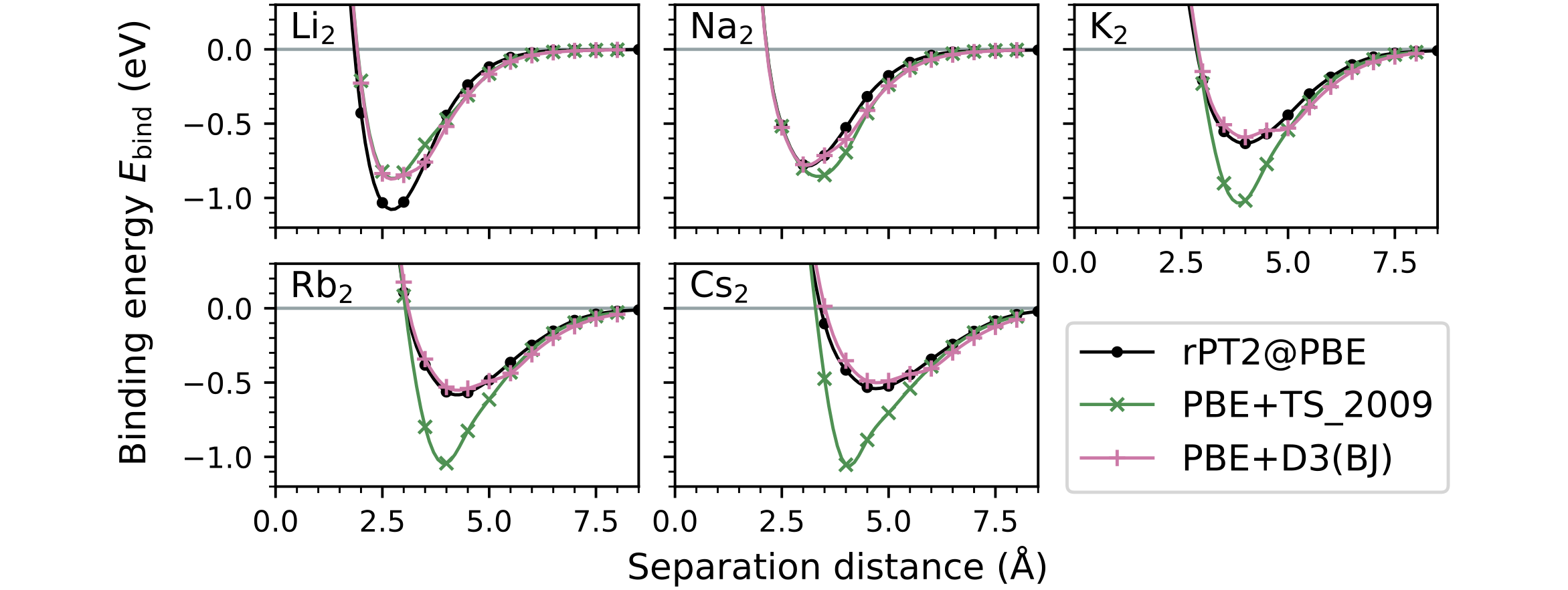}
\caption{{Binding energy for five alkali dimers calculated using PBE+TS\_2009 and PBE+D3(BJ), with the binding energy calculated with rPT2@PBE for reference. The PBE+TS\_2009 result exhibits a large overbinding, while the PBE+D3(BJ) result shows an unphysical second minimum at larger radius.}}
\label{fig:si-dimer-binding}
\end{figure*}

\twocolumngrid

\section{Reference values for covalent solids}
\label{sec:solid_reference}

To benchmark crystal structures obtained using different calculation methods, we compare against experimental unit cell volumes corrected for zero-point anharmonic expansion.
We use the model proposed in Ref.~\citenum{alchagirov2001energy} and derived in detail in Ref.~\citenum{hao2012lattice}, which is based on the Debye model and the Dugdale-MacDonald approximation for the Gr\"{u}neisen parameter \cite{dugdale1953thermal}.
In this model, the volume correction \emph{per atom} is given by
\[ \Delta v = \frac{9}{16} (B_1 - 1) \frac{k_B \Theta_D}{B_0} \]
where $\Delta v$ is the correction for zero-point expansion, $B_0$ is the bulk modulus, $B_1 = \left.\frac{d B_0}{dP}\right|_{v_0}$ is the pressure derivative of the bulk modulus at equilibrium volume $v_0$, and $\Theta_D$ is the Debye temperature.

Values for $B_0$, $B_1$, and $\Theta_D$ are obtained from experimental data and the resulting zero-point correction is subtracted from the experimental volume $v_0$.
This corrected volume is used as a reference.
The experimental values used for the alkali halides are listed in Table~\ref{tab:alk_expt_data}.
Unless noted otherwise, the data is obtained from Ref.~\citenum{sirdeshmukh2001alkali}.
For other materials, experimental reference data is listed in Table~\ref{tab:sc_expt_data}.
In cases where multiple data points are available, they are averaged.
In most cases, outliers are eliminated if the range of values is larger than \qty{10}{\percent} of the parameter value (see supporting information for details).
Given the relatively low values of the zero point corrections overall, we believe this approach is justified.
From these values, the zero-point corrections are calculated and listed in Table~\ref{tab:sc_vol}.
In the main text, $v_0 - \Delta v$ is used as the reference volume.

\begin{table}
\centering
\caption{Experimental reference data for alkali halides. Unless otherwise noted, data is obtained from tables 1.2, 2.11, 2.10, and 3.25 of Ref.~\citenum{sirdeshmukh2001alkali} for lattice constants $a_0$, $B_0$, $B_1$, and $\Theta_D$, respectively.
Lattice constants and elastic moduli are from room temperature data, and the Debye temperatures are from \qty{0}{\K}.
The volume correction per atom $\Delta v$ is calculated from these experimental results. Note that the rock salt structures have eight atoms per unit cell, while the \ce{CsCl} structures (\ce{CsCl}, \ce{CsBr}, \ce{CsI}) have two atoms per unit cell.
}
\label{tab:alk_expt_data}
\begin{ruledtabular}
\begin{tabular}{l d{1.5} d{2.1} d{1.2} d{3.1} d{1.3}}
Solid
& \multicolumn{1}{c}{$a_0$ (\unit{\angstrom})}
& \multicolumn{1}{c}{$B_0$ (\unit{\GPa})}
& \multicolumn{1}{c}{$B_1$}
& \multicolumn{1}{c}{$\Theta_D$ (\unit{\K})}
& \multicolumn{1}{c}{$\Delta v$ (\unit{\angstrom\cubed})} \\
\midrule
LiF  & 4.02620 & 67.1 & 5.21 & 733 & 0.357 \\
LiCl & 5.13988 & 29.7 & 5.42 & 429 & 0.496 \\
LiBr & 5.501   & 23.8 & 5.39 & 274 & 0.393 \\
LiI  & 6.012   & 17.1 & 5.79 & 210 & 0.457 \\
NaF  & 4.6329  & 46.5 & 5.18 & 492 & 0.343 \\
NaCl & 5.64009 & 24.0 & 5.27 & 320.8 & 0.443 \\
NaBr & 5.97299 & 19.9 & 5.29 & 224 & 0.375 \\
NaI  & 6.4728  & 15.1 & 5.40 & 167.5 & 0.379 \\
KF   & 5.344   & 30.5 & 5.26 & 332.8 & 0.361 \\
KCl  & 6.29294 & 17.4 & 5.34 & 236.1 & 0.457 \\
KBr  & 6.5982  & 14.8 & 5.38 & 172 & 0.395 \\
KI   & 7.0655  & 17.1 & 5.47 & 130.8 & 0.266 \\
RbF  & 5.6516  & 26.2 & 5.57 & 212\footnote{Room temperature value used} & 0.287 \\
RbCl & 6.5898  & 15.6 & 5.48 & 168.8 & 0.376 \\
RbBr & 6.8908  & 13.0 & 5.45 & 136.3 & 0.362 \\
RbI  & 7.3466  & 10.5 & 5.44 & 107.8 & 0.354 \\
%
CsCl & 4.1200  & 18.0 & 5.76 & 168 & 0.345 \\
CsBr & 4.2953  & 15.9 & 5.71 & 149.5 & 0.344 \\
CsI  & 4.5667  & 12.8 & 5.65 & 126.2 & 0.356 \\
\end{tabular}
\end{ruledtabular}
\end{table}

\begin{table*}
\centering
\caption{Experimental values of $B_0$, $B_1$, and $\Theta_D$. These values are an average if multiple data points are available.}
\label{tab:sc_expt_data}
\begin{ruledtabular}
\begin{tabular}{lc d{3.1} c d{1.1} c d{4.1} c}
Solid & Struct.
& \multicolumn{1}{c}{$B_0$ (\unit{\GPa})} & Ref.
& \multicolumn{1}{c}{$B_1$} & Ref.
& \multicolumn{1}{c}{$\Theta_D$ (\unit{\K})} & Ref. \\
\midrule
MgO & RS & 162.7 & \citenum{LandoltBornstein2017} & 4.0 & \citenum{LandoltBornstein2017} & 948 & \citenum{LandoltBornstein1999, Verma1975} \\
MgS & RS & 81 & \citenum{Peiris1994} & 3.6 & \citenum{Peiris1994} & 460 & \citenum{Baldwin1964} \\
C & Dia & 446 & \citenum{Occelli2003} & 3.0 & \citenum{Occelli2003} & 2219 & \citenum{Desnoyehs1958} \\
SiC & ZB & 233 & \citenum{LandoltBornstein2001} & 4.1 & \citenum{LandoltBornstein2001} & 1094 & \citenum{Slack1973,Siethoff1995} \\
PbS & RS & 57.5 & \citenum{LandoltBornstein1998, Bhagavantam1951} & 6.3 & \citenum{Peresada1976} & 220.1 & \citenum{LandoltBornstein1998, Knight2022, Verma1975} \\
PbSe & RS & 49.0 & \citenum{Knight2022, Li2014, Wang2015, Lippmann1971} & 5.4 & \citenum{Knight2022} & 154.5 & \citenum{LandoltBornstein1998, Knight2022} \\
PbTe & RS & 40.4 & \citenum{LandoltBornstein1998, Houston1968} & 5.2 & \citenum{Miller1981} & 127.5 & \citenum{Knight2022, Parkinson1954} \\
AlN & WUR & 201.3 & \citenum{LandoltBornstein2001} & 6.0 & \citenum{LandoltBornstein2001} & 842.7 & \citenum{Nipko1998aln, Meng1994} \\
AlAs & ZB & 74. & \citenum{LandoltBornstein2001} & 5. & \citenum{LandoltBornstein2001} & 415. & \citenum{LandoltBornstein2002} \\
AlSb & ZB & 56. & \citenum{LandoltBornstein2001} & 4.4 & \citenum{LandoltBornstein2001} & 293.4 & \citenum{LandoltBornstein2002, Verma1975} \\
GaN & WUR & 213.1 & \citenum{LandoltBornstein2001, Polian1996} & 3.8 & \citenum{LandoltBornstein2001} & 617.4 & \citenum{Nipko1998gan, Danilchenko2006, Passler2009} \\
GaP & ZB & 87.0 & \citenum{LandoltBornstein2001} & 4.6 & \citenum{LandoltBornstein2001} & 448.2 & \citenum{LandoltBornstein2002, Verma1975, Steigmeier1963, Boyle1975, Passler2009} \\
GaAs & ZB & 75.4 & \citenum{LandoltBornstein2001} & 4.5 & \citenum{LandoltBornstein2001} & 344.5 & \citenum{LandoltBornstein2002, Verma1975, Steigmeier1963, Passler2009} \\
GaSb & ZB & 55.4 & \citenum{LandoltBornstein2001} & 6. & \citenum{LandoltBornstein2001} & 267.6 & \citenum{LandoltBornstein2002, Steigmeier1963, Verma1975, Boyle1975, Holste1972, Passler2009} \\
InN & WUR & 128.3 & \citenum{LandoltBornstein2001, Yao2010} & 12.7 & \citenum{LandoltBornstein2001, Yao2010} & 370 & \citenum{Davydov1999} \\
InP & ZB & 73.4 & \citenum{LandoltBornstein2001} & 4.4 & \citenum{LandoltBornstein2001} & 309.1 & \citenum{LandoltBornstein2002, Verma1975, Steigmeier1963, Passler2009} \\
InAs & ZB & 58.6 & \citenum{LandoltBornstein2001} & 6.8 & \citenum{LandoltBornstein2001} & 253.5 & \citenum{LandoltBornstein2002, Verma1975, Steigmeier1963, Holste1972, Passler2009} \\
InSb & ZB & 45.7 & \citenum{LandoltBornstein2001} & 4.9 & \citenum{LandoltBornstein2001} & 204.9 & \citenum{LandoltBornstein2002, Steigmeier1963, Verma1975, Holste1972, Passler2009} \\
ZnO & WUR & 143.7 & \citenum{LandoltBornstein2013} & 3.6 & \citenum{Desgreniers1998} & 417.5 & \citenum{LandoltBornstein1999, Passler2009} \\
ZnS & ZB & 76.6 & \citenum{LandoltBornstein1999} & 4.9 & \citenum{LandoltBornstein1999} & 342.1 & \citenum{Collins1980, Verma1975, Passler2009, Birch1975} \\
ZnS & WUR & 76.5 & \citenum{Chang1973, Cline1967} & 4.4 & \citenum{Chang1973} & 351 & \citenum{Cline1967} \\
ZnSe & ZB & 62.0 & \citenum{LandoltBornstein2013, Lee1970} & 5.1 & \citenum{LandoltBornstein2013, Lee1970} & 272.7 & \citenum{LandoltBornstein1999, Collins1980, Passler2009} \\
ZnTe & ZB & 49.4 & \citenum{LandoltBornstein1999} & 4.9 & \citenum{LandoltBornstein1999} & 212.2 & \citenum{LandoltBornstein2009, Collins1980, Verma1975} \\
CdTe & ZB & 43.6 & \citenum{LandoltBornstein1999} & 5.1 & \citenum{LandoltBornstein1999} & 160.7 & \citenum{Birch1975, Collins1980, Verma1975, Passler2009} \\
HgSe & ZB & 50.1 & \citenum{LandoltBornstein1999} & 2.6 & \citenum{LandoltBornstein1999} & 148.0 & \citenum{LandoltBornstein1999, Verma1975, Lehoczky1969} \\
HgTe & ZB & 42.5 & \citenum{LandoltBornstein1999} & 3.8 & \citenum{LandoltBornstein1999} & 142.2 & \citenum{LandoltBornstein1999, Verma1975, Collins1980} \\
\end{tabular}
\end{ruledtabular}
\end{table*}

\begin{table}
\centering
\caption{Experimental volume per-atom $v_0$ from Ref.~\citenum{villars1985pearson} and the zero-point correction $\Delta v$ calculated using data from Table~\ref{tab:sc_expt_data}.}
\label{tab:sc_vol}
\begin{ruledtabular}
\begin{tabular}{lc d{2.2} d{1.2} d{1.2}}
Solid & Struct. 
& \multicolumn{1}{c}{$v_0$ (\unit{\angstrom\cubed})}
& \multicolumn{1}{c}{$\Delta v$ (\unit{\angstrom\cubed})}
& \multicolumn{1}{c}{$\Delta v / v_0$ (\unit{\percent})}\\
\midrule
MgO & RS & 9.36 & 0.14 & 1.47 \\
MgS & RS & 17.49 & 0.11 & 0.64 \\
C & Dia & 5.67 & 0.08 & 1.36 \\
SiC & ZB & 10.35 & 0.11 & 1.10 \\
PbS & RS & 26.15 & 0.16 & 0.60 \\
PbSe & RS & 28.74 & 0.11 & 0.37 \\
PbTe & RS & 33.70 & 0.10 & 0.30 \\
AlN & WUR & 10.43 & 0.16 & 1.56 \\
AlAs & ZB & 22.67 & 0.17 & 0.77 \\
AlSb & ZB & 28.87 & 0.14 & 0.48 \\
GaN & WUR & 11.43 & 0.06 & 0.56 \\
GaP & ZB & 20.24 & 0.15 & 0.72 \\
GaAs & ZB & 22.58 & 0.12 & 0.54 \\
GaSb & ZB & 28.32 & 0.19 & 0.66 \\
InN & WUR & 15.48 & 0.26 & 1.69 \\
InP & ZB & 25.27 & 0.11 & 0.45 \\
InAs & ZB & 27.80 & 0.19 & 0.70 \\
InSb & ZB & 34.00 & 0.14 & 0.40 \\
ZnO & WUR & 11.83 & 0.06 & 0.50 \\
ZnS & ZB & 19.80 & 0.14 & 0.69 \\
ZnS & WUR & 19.81 & 0.12 & 0.61 \\
ZnSe & ZB & 22.75 & 0.14 & 0.62 \\
ZnTe & ZB & 28.41 & 0.13 & 0.45 \\
CdTe & ZB & 33.92 & 0.12 & 0.35 \\
HgSe & ZB & 28.17 & 0.04 & 0.13 \\
HgTe & ZB & 33.68 & 0.07 & 0.21 \\
\end{tabular}
\end{ruledtabular}
\end{table}

\clearpage
\onecolumngrid

\section{Perovskite structure data}
\label{sec:perov_data}

\begin{table*}[hp]
\centering
\caption{The experimental and predicted lattice parameters (lattice vectors $a$,$b$, and $c$; lattice angles $\alpha$, $\beta$, and $\gamma$) of \ce{CsPbCl3}, \ce{CsPbBr3}, and \ce{CsPbI3}. All calculations use the 40 atom cell.}
\label{tab:perovskite_lattice_parameters}
\begin{ruledtabular}
\begin{tabular}{l| c | d{2}d{2}d{2} d{2}d{2}d{2}}
System & Method
& \multicolumn{1}{c}{$a$ (\unit{\angstrom})}
& \multicolumn{1}{c}{$b$ (\unit{\angstrom})}
& \multicolumn{1}{c}{$c$ (\unit{\angstrom})}
& \multicolumn{1}{c}{$\alpha$ (\unit{\degree})}
& \multicolumn{1}{c}{$\beta$ (\unit{\degree})}
& \multicolumn{1}{c}{$\gamma$ (\unit{\degree})} \\
\midrule
\multirow{8}{4em}{\ce{CsPbCl3}}
 & Experiment\cite{muscarella2023which} (\qty{90}{\K})
                  & 11.11 & 11.19 & 11.11 & 90.0 & 90.9 & 90.0 \\
 & PBE            & 11.36 & 11.37 & 11.36 & 90.0 & 91.2 & 90.0 \\
 & PBE+TS\_2009   & 10.80 & 11.40 & 10.74 & 90.0 & 93.0 & 90.0 \\
 & PBE+TS\_2018   & 11.22 & 11.32 & 11.22 & 90.0 & 93.5 & 90.0 \\
 & PBE+TS\_alkali & 11.32 & 11.28 & 11.32 & 90.0 & 90.3 & 90.0 \\
 & PBE+D3(BJ)     & 11.22 & 11.27 & 11.22 & 90.0 & 91.3 & 90.0  \\
 & B86bPBE+XDM    & 11.10 & 11.22 & 11.10 & 90.0 & 91.3 & 90.0 \\ 
 & PBE+MBD-NL    & 11.13 & 11.23 & 11.13 & 90.0 & 94.0 & 90.0 \\
\midrule
\multirow{8}{4em}{\ce{CsPbBr3}}
 & Experiment\cite{muscarella2023which} (\qty{90}{\K})
                  & 11.58 & 11.68 & 11.58 & 90.0 & 91.4 & 90.0 \\
 & PBE            & 11.88 & 11.90 & 11.88 & 90.0 & 91.6 & 90.0 \\
 & PBE+TS\_2009   & 11.38 & 11.86 & 11.38 & 90.0 & 94.9 & 90.0 \\
 & PBE+TS\_2018   & 11.75 & 11.79 & 11.75 & 90.0 & 92.4 & 90.0 \\
 & PBE+TS\_alkali & 11.81 & 11.78 & 11.81 & 90.0 & 90.3 & 90.0 \\
 & PBE+D3(BJ)     & 11.69 & 11.78 & 11.69 & 90.0 & 92.0 & 90.0 \\
 & B86bPBE+XDM    & 11.54 & 11.75 & 11.54 & 90.0 & 92.5 & 90.0 \\
 & PBE+MBD-NL    & 11.60 & 11.74 & 11.60 & 90.0 & 95.1 & 90.0 \\
\midrule
\multirow{8}{4em}{\ce{CsPbI3}}
 & Experiment\cite{straus2020understanding} (\qty{100}{\K})
                  & 12.24 & 12.41 & 12.24 & 90.0 & 94.1 & 90.0 \\
 & PBE            & 12.63 & 12.65 & 12.63 & 90.0 & 92.3 & 90.0 \\
 & PBE+TS\_2009   & 12.11 & 12.58 & 12.11 & 90.0 & 97.0 & 90.0 \\
 & PBE+TS\_2018   & 12.48 & 12.45 & 12.48 & 90.0 & 91.4 & 90.0 \\
 & PBE+TS\_alkali & 12.46 & 12.48 & 12.46 & 90.0 & 90.9 & 90.0 \\
 & PBE+D3(BJ)     & 12.35 & 12.51 & 12.35 & 90.0 & 93.6 & 90.0 \\
 & B86bPBE+XDM    & 12.13 & 12.49 & 12.13 & 90.0 & 96.1 & 90.0 \\
 & PBE+MBD-NL    & 12.25 & 12.44 & 12.25 & 90.0 & 96.7 & 90.0 \\
\end{tabular}
\end{ruledtabular}
\end{table*}

\begin{table}[hp]
\centering
\caption{Error in the cell volumes ($V$), the cell anisotropy ($b/a$), and the $\beta$ angle of \ce{CsPbCl3}, \ce{CsPbBr3}, and \ce{CsPbI3} in comparison with experimental structures. Same references and experimental temperatures as in Table \ref{tab:perovskite_lattice_parameters}.}
\label{tab:volume_and_anisotropy}
\begin{ruledtabular}
\begin{tabular}{l|c| d{2}d{2}d{2}}
System & Method
& \multicolumn{1}{c}{$V$ error (\unit{\percent})}
& \multicolumn{1}{c}{$b/a$ error (\unit{\percent})} 
& \multicolumn{1}{c}{$\beta$ error (\unit{\degree})} \\
\midrule
\multirow{7}{4em}{\ce{CsPbCl3}}
 & PBE            & 6.43 & -0.63 & 0.35 \\
 & PBE+TS\_2009   & -4.34 & 4.78 & 2.15 \\
 & PBE+TS\_2018   & 3.10 & 0.09 & 2.57 \\
 & PBE+TS\_alkali & 4.70 & -1.06 & -0.56 \\
 & PBE+D3(BJ)     & 2.73 & -0.22 & 0.38 \\
 & B86bPBE+XDM    & 0.26 & 0.35 & 0.37\\
 & PBE+MBD-NL    & 0.53 & 0.24 & 3.09 \\
\midrule
\multirow{7}{4em}{\ce{CsPbBr3}}
 & PBE            & 7.11 & -0.69 & 0.24 \\
 & PBE+TS\_2009   & -2.34 & 3.37 & 3.53 \\
 & PBE+TS\_2018   & 3.78 & -0.48 & 0.96 \\
 & PBE+TS\_alkali & 4.86 & -1.11 & -1.11 \\
 & PBE+D3(BJ)     & 2.64 & -0.03 & 0.58 \\
 & B86bPBE+XDM    & -0.12 & 0.98 & 1.15 \\
 & PBE+MBD-NL    & 0.33 & 0.40 & 3.66 \\
\midrule
\multirow{7}{4em}{\ce{CsPbI3}}
 & PBE            & 8.67 & -1.18 & -1.80 \\
 & PBE+TS\_2009   & -1.17 & 2.43 & 2.94 \\
 & PBE+TS\_2018   & 4.47 & -1.62 & -2.73 \\
 & PBE+TS\_alkali & 4.35 & -1.33 & -3.19 \\
 & PBE+D3(BJ)     & 2.61 & -0.08 & -0.46 \\
 & B86bPBE+XDM    & -1.50 & 1.50 & 2.01 \\
 & PBE+MBD-NL    & -0.05 & 0.08 & 2.59 \\
\end{tabular}
\end{ruledtabular}
\end{table}

\clearpage

\section{Timings for vdW methods}
\label{sec:timings}

\begin{table}[h]
\centering
\caption{{Timings for a single point energy calculation of a 40-atom \ce{CsPbBr3} structure using the FHI-aims code with the ``tight'' numerical defaults and associated basis set, a \numproduct{4x4x4} k-point grid, and a \texttt{radial\_multiplier} of 4.
The calculation was performed with \num{192} MPI tasks on a single dual-socket AMD EPYC 9654 node of the Sycamore Cluster at the University of North Carolina at Chapel Hill (UNC).}
{Note that the van der Waals corrections in question are not executed in parallel. They were used in the context of their existing implementations in FHI-aims and in libMBD\cite{blum2009ab, abbott2026roadmap, hermann2023libmbd}.
Optimizing their performance was not a focus of this work---the provided timings merely show that they are essentially negligible compared to the remaining SCF calculation, even when that remaining SCF calculation is parallelized over \num{192} cores.}
}
\label{tab:timings}
\begin{ruledtabular}
\begin{tabular}{ld{1}d{1}d{1}}
Method 
& \multicolumn{1}{c}{SCF init. (s)}
& \multicolumn{1}{c}{vdW calc. (s)}
& \multicolumn{1}{c}{Total time (s)} \\
\midrule
PBE             & 8.1 & {-} & 109.3 \\
PBE+TS\_2009    & 8.1 & 2.3 & 110.6 \\
PBE+TS\_2018    & 8.1 & 2.4 & 110.9 \\
PBE+TS\_alkali  & 8.1 & 2.3 & 111.0 \\
PBE+D3(BJ)      & 8.2 & 0.7 & 109.5 \\
B86bPBE+XDM     & 8.3 & 8.2 & 128.7 \\
PBE+MBD-NL      & 8.4 & 6.1 & 115.7 \\
\end{tabular}
\end{ruledtabular}
\end{table}

\clearpage
\twocolumngrid

\bibliography{references}

\end{document}